\documentclass[11pt,letterpaper,aps,preprint,groupedaddress,showpacs,amsmath,amssymb]{revtex4-1}
\usepackage[margin=1.75cm]{geometry}
\usepackage{setspace}
\DeclareRobustCommand{\rchi}{{\mathpalette\irchi\relax}}
\newcommand{\irchi}[2]{\raisebox{\depth}{$#1\chi$}}
\usepackage{graphicx}
\usepackage{dcolumn}
\usepackage{bm}
\usepackage{setspace}

\begin{document}

\title{Single-particle and collective excitations in quantum wires comprised of vertically stacked quantum dots:
Finite magnetic field\\}
\author{Manvir S. Kushwaha$^1$}

\address
{\centerline {Department of Physics and Astronomy, Rice University, P.O. Box 1892, Houston, TX 77251, USA}\\
$^1$ Corresponding Author: manvir@rice.edu}
\date{\today}

\begin{abstract}
A theoretical investigation has been made of the magnetoplasmon excitations in a quasi-one-dimensional electron system comprised of vertically stacked, self-assembled InAs/GaAs quantum dots. The smaller length scales involved in the experiments impel us to consider a perfectly periodic system of two-dimensionally confined InAs quantum dot layers separated by GaAs spacers. Subsequent system is subjected to a two-dimensional confining (harmonic) potential in the x-y plane and an applied magnetic field (B) in the symmetric gauge. This scheme defines virtually a system of quantum wire comprised of vertically stacked quantum dots (VSQD). We derive and discuss the Dyson equation, the generalized (nonlocal and dynamic) dielectric function, and the inverse dielectric function for investigating the single-particle and collective (magnetoplasmon) excitations within the framework of (full) random-phase approximation (RPA). As an application, we study the influence of the confinement potential and the magnetic field on the component eigenfunctions, the density of states (DOS), the Fermi energy, the collective excitations, and the inverse dielectric functions. How the B-dependence of DOS validate the VSQD mimicking the realistic quantum wires, the Fermi energy oscillates as a function of the Bloch vector, the intersubband single-particle continuum bifurcates at the origin, a collective excitation emerges and propagates within the gap of the split single-particle continuum, and the alteration in the well- and barrier-widths allows to customize the excitation spectrum in the desired energy range are some of the remarkable features of this investigation. These findings demonstrate, for the very first time, the significance of investigating the system of VSQD subjected to a quantizing magnetic field. Given the edge over the planar quantum dots and the foreseen applications in the single-electron devices and quantum computation, investigating the system
of VSQD is deemed vital. The results suggest exploiting magnetoplasmon qubits to be a potential option for implementing the solemn idea of quantum state transfer in devising quantum gates for the quantum computation and quantum communication networks.
\end{abstract}
\pacs{73.21.-b, 73.63.-b, 75.47.-m, 78.20.Ls}
\maketitle



\section{Introduction}

A closer look at the literature of condensed-matter physics of the past three decades reveals the impact, the
legacy, and the ongoing influence of the man-made, quasi-n-dimensional, semiconducting heterostructures -- $n$
($=2$, 1, or 0) being the degree of freedom -- on the current (and the near future) science and technology.
Thanks to the wishful advancements in the nanofabrication technology and the electron lithography, these
quantum structures have made possible the emergence of much of the exotic -- fundamental and applied -- physics
transpired during this period. The continued interest in the physics and fabrication of such tailor-made
nanostructures is triggered by the world-wide drive to develop exciting high-speed, low-threshold devices that
are small enough, sharp enough, or uniform enough to behave the way theory says they should. Fabricated mostly
from III-V compounds and fashionably becoming known as the quantum wells (with $n=2$), quantum wires
(with $n=1$), and quantum dots (with $n=0$), these nanostructures represent the systems in which the charge
carriers exposed to external electric and/or magnetic fields can (and do) manifest unusual quantal effects that
strongly alter their behavior characteristics [1].

It is beyond dispute that the discovery of quantum Hall effects -- both integer and fractional -- had spurred the
research interest in semiconducting systems of lower dimensions, which keeps electrifying a vast majority of
condensed matter physicists [1]. The emergence of graphene in 2004 and its device potential quickly stimulated
researchers to inquire into other materials of the same descent, which include germanene and silicene. The
optimism in the research on the ...enes has also prompted several research groups to explore layered materials
such as nitrides, oxides, selenides, sulfides, and tellurides, which have gradually joined the long list of
largely exploited arsenides, antimonides, and phosphides. Consequently, the making of nanostructures like
nanoribbons, nanowires, nanoballs, nanorods, nanogrids, nanopillars, nanosnakes, nanoonions, ...etc. keeps
researchers captivated by the miniaturization. However, nothing compares to the total confinement of quantum dots,
which have many features in common with the real atoms. Nevertheless, the pivotal difference is mesmerizing:
numerous electronic, optical, and transport properties in quantum dots are tunable – unlike in the real atoms.
The tunability allows the wishful control and manipulation of the charge carriers in the system. This makes the
quantum dots, especially the self-assembled quantum dots, the real gems for devising nanodevices, which forge the
foundation of future nanoscience and nanotechnology.

Smaller length scales (of a few nanometers) involved in the experimental setup prompt us to consider a perfectly
periodic system of two-dimensionally confined InAs quantum dot layers separated by GaAs spacers. Given the adequate
lateral confinement and vertical coupling, the resultant system can be made to mimic the quantum wires comprised of
quantum dots. The readers are reminded of certain facts which motivate us to exploit the periodic system of VSQD
for the device designing. These are: (i) Sakaki's meticulous diagnosis that led him to justify the designing of
such heterostructures in which the optical phonon scattering can be essentially eliminated [2]; (ii) the methodical justification for the use of Fermi-liquid-like theories for describing the realistic quantum wires [3]; (iii) the
strain due to lattice mismatch at the interfaces, which provides impetus for the growth of self-assembled quantum
dots in the system [4, 5]; (iv) the wishful tunability of the coupling strength, which not merely paves the way to
interesting applications in quantum computation and spintronics, it also allows exploring new fundamental physics
of steering spins and controlling electron g factors [6]; and (v) the robust coupling strength along the growth
axis rendering a quasi-one-dimensional (Q1D) system devised of quasi-zero-dimensional (Q0D) systems. Here item (v)
fulfills the quest for {\em reversing the trend}, i.e., fabricating the systems of higher dimensions [1]:
0 to 1, 1 to 2, and 2 to 3.

Research interest in the system of VSQD fabricated out of InAs islands separated by GaAs spacers along the growth
axis was triggered by early experiments in the mid-1990s [4-6]. The foreseen device potential of the VSQD system
immediately prompted many research groups embarking on the wide-ranging investigations. Consequently, diverse
experimental [7-56] and theoretical [57-76] works started dealing with electronic, optical, and transport
phenomena, which persuaded authors to envision diverse solid-state devices. The early focus remained primarily on
the quantum-dot molecules (QDM) made up of the pairs of VSQD separated by thin barriers. The reason was their
significance in realizing the short-range quantum-state transfers. As a matter of fact, this concept proved to be
consequential for the future quantum communication networks. In most cases, it is imperative to stack multiple VSQD
in order to allow a larger flux of emitted or absorbed photons. Their aerial number density depends on the specific
use, however.

Initial experimental as well as theoretical focus -- based on the shape, size, and setup of the VSQD -- had mainly
been restricted to a single isolated qubit, whereas the real fuss in quantum computation calls for creating and
controlling entanglement of multiple qubits. The basic notion in realizing the quantum gates, which are the salient
units of quantum communication networks, rests in the consistent manipulation of exciton qubits. Interestingly, the
role of an applied magnetic field relevant to the (nuclear or electron) polarization does not seem to be cherished
enough in the VSQD. In spite of the exotic role played by the magnetoplasmon excitations in the electronic, optical,
and transport phenomena in diverse quantum nanostructures [1], the literature is still devoid of any report on
proposing novel magnetoplasmon qubits (MPQ), which offer a greater speed advantage over the ritual exciton qubits
in the VSQD. This article tends to fill that void.


The rest of the article is organized as follows. In Sec. II, we present the methodological framework leading to
the derivation of the nonlocal, dynamic, dielectric function (within the full random-phase approximation (RPA)
[77]), which is further scrutinized to fully address the solution of the problem and the relevant physics. In
Sec. III, we discuss several illustrative examples of, for instance, the density of states, the Fermi energy,
Coulomb interaction, the single-particle and collective excitation spectrum, ...etc. There, we also examine the
effects of an applied magnetic field and the influence of alteration in the well- and barrier-thicknesses on the
excitation spectrum. In addition, we call attention to the importance of studying inverse dielectric function in
connection with the quantum transport phenomena in the system [78]. Finally, in Sec. IV, we summarize our findings
and highlight some interesting features worth adding to the problem.

\section{Theoretical Framework}

\subsection{Eigenfunctions and eigenenergies}

We consider a periodic system of quasi-two-dimensional InAs islands of thickness $a$ separated by GaAs spacer layers
of thickness $b$. The InAs islands are constrained by a two-dimensional confining (harmonic) potential of the form
of $V(x)=\frac{1}{2}m^*\omega_0^2 (x^2+y^2)$ in the x-y plane and subjected to an applied magnetic field (B) in the
symmetric gauge specified by the vector potential ${\bf A}=\frac{1}{2}\,B$ (-$y$, $x$). This choice of the vector
potential preserves the gauge invariance brought about by the so-called gauge transformation that requires the
electromagnetic fields ${\bf E}$ and ${\bf B}$ to remain the same. We assume the growth axis to be under a robust
confining potential, $V_c (z)$, yielding strong coupling between InAs layers. The small length scales and strong
coupling make the final system of VSQD (see Fig. 1) mimic a realistic quantum wire with a well-defined {\em linear}
charge density and hence legitimize the tight-binding approximation (TBA) [62]. The moderate tunneling (in the
polarizability function) is accounted for with the energy dispersion being sinusoidal. The effective quantum wire is characterized by the single-particle [of charge $-e$, with $e>0$] Hamiltonian
\begin{figure}[htbp]
\includegraphics*[width=8cm,height=9cm]{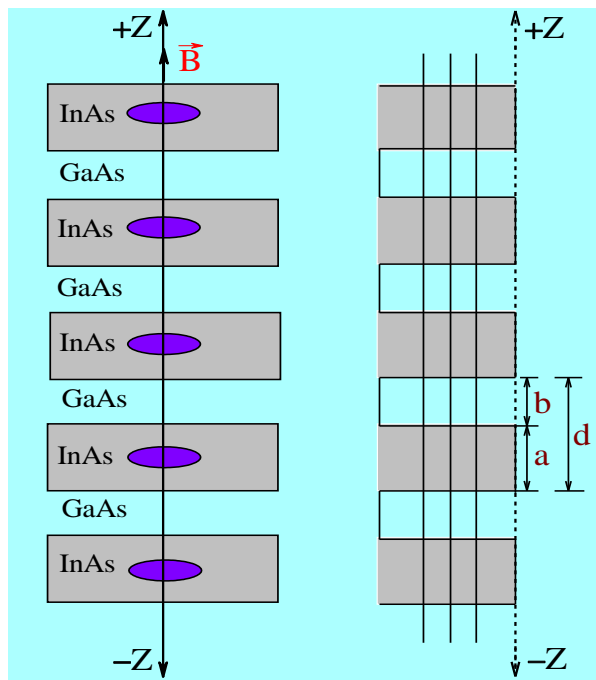}
\caption{(Color online) Schematics of the quantum wire made up of an infinitely periodic system of InAs islands
separated by GaAs spacer layers (left panel). The grey (purple) area in the left panel refers to practically
active region before (after) constraining the InAs islands with harmonic confinement in the x-y plane and the
magnetic field in the symmetric gauge. The right panel shows the Kronig-Penney periodic-potential simulation along
the growth direction. Here $a$ ($b$) is the well (barrier) width and $d=a+b$ is the period of the resultant system
making up the quantum wire. (After Kushwaha, Ref. 85).}
\label{fig1}
\end{figure}

\begin{align}
H=\frac{1}{2m^*}\,\Big[\Big(p_x + \frac{e}{c}A_x\Big)^2 + \Big(p_y + \frac{e}{c}A_y\Big)^2\Big]
                + \frac{1}{2}\,m^* \,\omega_o^2 \,(x^2 + y^2)\, + \frac{1}{2m^*}\,p_z^2  + V_c(z),
\end{align}
in the Coulomb gauge, where $\omega_o$ is the characteristic frequency of the bare harmonic oscillator. The strain
effect, which arises due to the lattice-mismatched hosts of the system and matters most for the phonon excitations,
is virtually nonexistent on the collective excitations such as plasmons and magnetoplasmons.
That is why we choose to exclude the strain effect from our theoretical framework. We ignore the spin-orbit
interactions and the Zeeman energy for the sake of simplicity. Since the use of cylindrical polar coordinates seems
to make more sense for the formal analysis, we cast the corresponding Schrodinger equation in the form
\begin{align}
\Big[-\frac{\hbar^2}{2m^*}\,\nabla^2_{\parallel}-i\,\frac{\hbar}{2}\,\omega_c\partial_{\theta}+
                \frac{1}{8}\,m^* \,\Omega^2\,r^2\Big]\Psi(r,\theta, z)
    - \Big[\frac{\hbar^2}{2m^*}\,\partial^2_z - V_c(z)\Big]\Psi(r,\theta, z)=E\Psi(r,\theta, z),
\end{align}
where $\partial_x \equiv \partial/{\partial x}$, $\nabla^2_{\parallel}=\big(\partial^2_r + \tfrac{1}{r}\partial_r + \tfrac{1}{r^2}\partial^2_{\theta}\big)$, $\omega_c=eB/(m^* c)$ is the cyclotron frequency, and
$\Omega=\sqrt{\omega^2_c + 4\omega^2_o}$ the effective characteristic frequency of the dressed harmonic oscillator.
Such a system as described above is solved by employing the method of separation of variables and formally
characterized by the eigenfunction
\begin{align}
\Psi(r, \theta, z)=\phi_n(r)\,\phi_m(\theta)\,\phi_k(z)\, ,
\end{align}
in the polar coordinates, where
\begin{align}
\phi_n(r)=\frac{1}{\ell_e}\,\sqrt{\frac{(n+\alpha)!}{n!(\alpha!)^2}}\,e^{-y/2}\,y^{\alpha/2}\,\Phi(-n,\alpha+1,y),\\
\phi_m(\theta)=\frac{1}{\sqrt{2\pi}}\,e^{im\theta}\, ,\\
\phi_k (z)=\frac{1}{\sqrt{N}}\,\sum_l\, e^{i k l d}\,\rchi_t (z - l d)\, ,
\end{align}
where $y=r^2/2\ell_e^2$, $n$ is the subband index, $\alpha=\vert m\vert$, $m$ being the orbital quantum number,
$d=a+b$ the period, $k$ the Bloch vector, $\ell_e =\sqrt{\hbar/m^*\Omega}$ the effective magnetic length,
$\Phi (...)$ the confluent hypergeometric function [79], and $\chi_{_{t}}(...)$ the Wannier function; and the
eigenenergy
\begin{align}
\epsilon_{nmk}=\!\big[n+\tfrac{1}{2}(\alpha+1)\big]\hbar\Omega\! + \! \tfrac{1}{2}m\hbar\omega_c\! + \!
                \epsilon_t\! - \!\tfrac{W_t}{2}\cos(kd),
\end{align}
where $\epsilon_t$ is the energy of the $t$-th miniband and $W_t$ is the band-width defined by
\begin{equation}
W_t = - 4 \int^{+a/2}_{-a/2} dz\,\, \rchi_{t} (z)\, V_{_0}\, \rchi_{t}(z-d),
\end{equation}
where we assume that the confining potential is a finite square well (Kronig-Penney potential) with a barrier
height $V_0$ and well-width $a$. Since $N$ (the number of quantum dot layers) is very large, the sum in Eq. (6)
can be written as integral according to the replacement rule: $\sum_k\, \rightarrow \,(N/L_{z})\, \int_{_{BZ}}\,
dk$. Eq. (6) represents the tight-binding constraint which hypothesizes a little overlap between the wave
functions of different sites -- with $t$ as the miniband index. There $\chi_{_{t}}(...)$, if normalized in the
length of the lattice (or, nearly enough, in infinite length), satisfies: $\int dz\,\chi^*_{_{t}}(z-nd)\,\chi_{_{t}}(z-ld)=\delta_{nl}$ and
$\int dz\,\phi^*_{t}(z)\,\phi_{t}(z)=1$. Here $L_z=Nd$ is the total crystal length along the growth direction.

\vspace{2.0cm}

\subsection{Generalized dielectric function}

We start with the general expression of the non-interacting single-particle density-density response function
(DDRF) $\chi^0 (...)$ given by [1]
\begin{equation}
\chi^{0} ({\bf r},{\bf r'};\omega)=\sum_{ij}\, \Lambda_{ij}\,\,
\psi^*_i ({\bf r})\,\psi_j ({\bf r})\,
\psi^*_j ({\bf r'})\,\psi_i ({\bf r'}),
\end{equation}
where ${\bf r}\equiv ({\bf{r}'_{\parallel}},z)$, with ${\bf{r}'_{\parallel}}\equiv (r, \theta)$; the composite index
$i,j\equiv k,n,t$; and $\Lambda_{ij}$ is defined as follows.

\begin{equation}
\Lambda_{ij}= 2\, \frac{f(\epsilon_i)-f(\epsilon_j)}{\epsilon_i-\epsilon_j+\hbar\omega^*},
\end{equation}
where $f(x)$ is the well-known Fermi distribution function. $\omega^*=\omega+i\gamma$ and small but nonzero $\gamma$
represents the adiabatic switching of the Coulomb interactions in the remote past. The factor of $2$ takes care of
the spin degeneracy. Next, we make use of the Kubo's correlation function to write the induced particle density
\begin{align}
n_{in}({\bf{r}}; \omega)
 &=\! \int d{\bf{r}'}\, \rchi^0({\bf{r}},{\bf{r}'},\omega) V_{tot}({\bf{r}'}; \omega)\\
 &=\! \int d{\bf{r}'}\, \rchi({\bf{r}},{\bf{r}'},\omega) V_{ex}({\bf{r}'}; \omega),
\end{align}
where $V_{tot}=V_{ex}+V_{in}$ is the total potential, with $V_{ex}$ ($V_{in}$) as the external (induced) potential.
$\rchi$ and $\rchi^o$ are, respectively, the total (or interacting) and single-particle DDRF and can be shown to be
related with each other through the integral Dyson equation [see, e.g., Fig. 2]
\begin{align}
\rchi({\bf{r}}, {\bf{r}'}; \omega)=
\rchi^0({\bf{r}}, {\bf{r}'}; \omega)+\int d{\bf{r}''}\int d{\bf{r}'''} 
\rchi^0({\bf{r}}, {\bf{r}''}; \omega)\,V_{ee}({\bf{r}''}, {\bf{r}'''})\,\rchi({\bf{r}'''}, {\bf{r}'}; \omega),
\end{align}
where $V_{ee}$(...) stands for the binary Coulombic  interactions in the direct space and is defined by
\begin{align}
V_{ee}({\bf r}, {\bf r'})
=\frac{e^2}{\epsilon_b}\,\frac{1}{\mid {\bf r}-{\bf r'} \mid}
=\frac{e^2}{\epsilon_b}\,\frac{1}{\mid ({\bf{r}_{\parallel}}-{\bf{r}'_{\parallel}})^2+(z-z')^2 \mid^{1/2}},
\end{align}
where $\epsilon_b$ the background dielectric constant of the system.
\begin{figure}[htbp]
\includegraphics*[width=9.9cm,height=5.0cm]{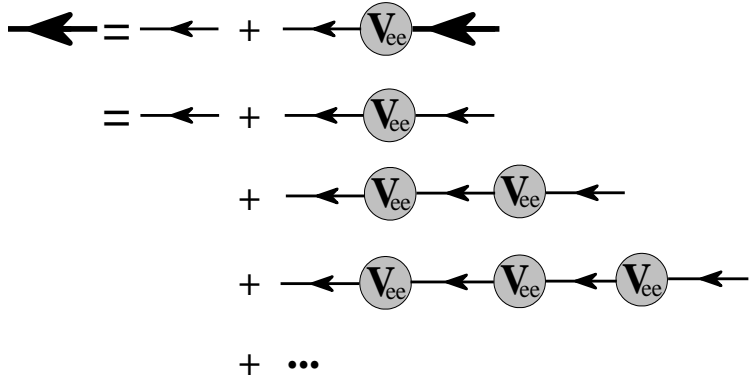}
\caption{A Feynman diagram for the Dyson equation: the thick [thin] line represents the total [single-particle] DDRF
$\rchi (...)$ [$\rchi^o (...)$] within the full RPA [77]. Here $V_{ee}$ represents the binary Coulombic interactions
in the direct space. The arrows indicate the transition from initial to final spatio-temporal position of the particle
in the interaction picture of the process.}
\label{fig2}
\end{figure}
Next, the induced potential in terms of the induced particle density is expressed as
\begin{align}
V_{in}({\bf{r}},\omega) = \int d{\bf{r}'}\,V_{ee} ({\bf{r}},{\bf{r}'})\, n_{in} ({\bf{r}'},\omega)
\end{align}
This equation -- with the aid of Eqs. (9) and (11) -- can be cast in the following form:
\begin{align}
V_{in}(\bf{r}, \omega)
=& \int d{\bf{r}'}\!\!\!\int d{\bf{r}''}\,V_{ee}({\bf{r}},{\bf{r}'})\,
                                \rchi^0 ({\bf{r}'},{\bf{r}''},\omega)\,V_{tot} ({\bf{r}''}; \omega)\nonumber\\
=& \sum_{i,j}\Lambda_{ij}\int d{\bf{r}'}\!\!\!\int d{\bf{r}''}\,V_{ee}({\bf{r}},{\bf{r}'})\,
                                                              \Psi^*_i ({\bf{r}'})\Psi_j({\bf{r}'})\nonumber\\
  &\times \Psi^*_j ({\bf{r}''})\Psi_i({\bf{r}''})\, V_{tot}({\bf{r}''}; \omega).
\end{align}
Making use of the explicit forms of the eigenfunctions in Eqs. (4)-(6) allows us to write Eq. (16) as
\begin{align}
V_{in}(\bf{r}_{\parallel}, z; \omega)=\frac{1}{N^2}&\sum_{nn'}\,\sum_{mm'}\,\sum_{tt'}\,\sum_{kk'}\,\sum_{ll'}
                         \,\Lambda_{\stackrel{nn'mm'}{tt'll'}}(k,k'; \omega)\nonumber\\
&\times\!\! \int \!d{{\bf{r}'}_{\parallel}}\!\!\int \! dz'\!\!\int \! d{{\bf{r}''}_{\parallel}}\!\!\int \! dz''
      \,V_{ee}({{\bf{r}}_{\parallel}},{{\bf{r}'}_{\parallel}}; z-z')\nonumber\\
&\times \, \phi^*_{nm}({\bf{r}'_{\parallel}})\,\phi_{n'm'}({\bf{r}'_{\parallel}})\,
           \phi^*_{n'm'}({\bf{r}''_{\parallel}})  \,\phi_{nm}({{\bf{r}''}_{\parallel}})\nonumber\\
&\times \, \scalebox{1.25}{$e^{-ikld}\,e^{ik'ld}\,e^{-ik'l'd}\,e^{ikl'd}$}\nonumber\\
&\times \, \scalebox{0.90}{$\rchi^*_{t} (z'\! -\! ld)\,\rchi_{t'} (z'\! -\! ld)\,\rchi^*_{t'} (z''\! -\! l'd)
           \rchi_{t} (z''\!\! -\! l'd)$}\nonumber\\
&\times \, V_{tot}({\bf{r}''_{\parallel}}, z''; \omega),
\end{align}
where $k'=k+q$, with $q$ as the momentum transfer. Next,  we (i) multiply both sides of this equation by $e^{-iq' z}$
and integrate with respect to $z$ and (ii) introduce , for convenience, a single Fourier component of $V_{tot}$(...)
to write $V_{tot}({\bf{r}}_{\parallel}, z; \omega)=e^{iq' z}\,V_{tot}({\bf{r}}_{\parallel}, q'; \omega)$. So, let us
first rewrite Eq. (17) in clearer terms before we open up the sums over $ll'$ and $kk'$. After rearranging a few terms,
Eq. (17) thus assumes the following form.
\begin{align}
V_{in}({\bf{r}}_{\parallel}, q'; \omega)=\frac{1}{N^2}&\sum_{nn'}\,\sum_{mm'}\,\sum_{tt'}\,\sum_{kk'}\,\sum_{ll'}
 \,\Lambda_{\stackrel{nn'mm'}{tt'll'}}(k,k'; \omega)\nonumber\\
&\times \int \!d{\bf{r}}'_{\parallel}\,
        V_{ee}({\bf{r}}_{\parallel}-{\bf{r'}}_{\parallel}; q')\,
        \phi^*_{nm}({\bf{r}'}_{\parallel})\,\phi_{n'm'}({\bf{r}'}_{\parallel})\nonumber\\
&\times \int \! d{{\bf{r}''}_{\parallel}}\,
        \phi^*_{n'm'}({\bf{r}''}_{\parallel})\,V_{tot}({\bf{r}''}_{\parallel}, z''; \omega)\,
        \phi_{nm}({\bf{r}''}_{\parallel})\nonumber\\
&\times \, \scalebox{1.25}{$e^{iqld}\,e^{-iql'd}\,e^{-iq'ld}\,e^{iq'l'd}$}\nonumber\\
&\times \int \! dz'\,  e^{-iq'(z'-ld)} \rchi^*_{t} (z'\! -\! ld)\,\rchi_{t'} (z'\! -\! ld)\nonumber\\
&\times \int \! dz''\, e^{iq'(z''-l'd)} \rchi_{t} (z''\! -\! l'd)\rchi^*_{t'} (z''\!\! -\! l'd),
\end{align}
where the Fourier-transformed Coulombic interaction
\begin{align}
V_{ee}({\bf{r}}_{\parallel}-{\bf{r'}}_{\parallel};q')=\frac{2e^2}{\epsilon_b}\,
                                           K_0\Big(q'\vert{\bf{r}}_{\parallel}-{\bf{r'}}_{\parallel}\vert\Big),
\end{align}
where $K_0$(...) is the zeroth-order modified Bessel function of the second kind. Since the last two integrals in
Eq. (18) can very well be jointly written as
\begin{align}
S_{tt'}(q)=\Big \vert \int dz\,  e^{-iqz} \rchi^*_{t}(z)\,\rchi_{t'}(z)\Big \vert^2,
\end{align}
we obtain, from the sum over $l'$,
\begin{align}
\frac{1}{N}\sum_{l'}\,e^{i(q'-q)(l'-l)d}
=\frac{1}{Nd}\,\int d(l'd)\,e^{i(q'-q)(l'-l)d}
=\frac{1}{L_z}\,2\pi\delta(q-q'),
\end{align}
\noindent and the remaining sum over $l$ simplifies to $\frac{1}{N}\sum_{l}\,1 = 1$. As a result, the sum over $k'$
solves as follows:
\begin{align}
\frac{2\pi}{L_z}\,\sum_{k'}\Lambda_{\stackrel{nn'mm'}{tt'}}(k, k'; \omega)\,\delta(q-q')
&=\int dk'\, \Lambda_{\stackrel{nn'mm'}{tt'}}(k, k'; \omega)\,\delta(k'-k-q')\nonumber\\
&=\Lambda_{\stackrel{nn'mm'}{tt'}}(k, k'=k+q'; \omega).
\end{align}
Next, we make use of the substitution
\begin{align}
\Pi_{nn'mm'tt'}(k, k'; \omega)=\sum_k\,\Lambda_{\stackrel{nn'mm'}{tt'}}(k, k'=k+q'; \omega),
\end{align}
and replace $q'$ safely by $q$ everywhere. As such, Eq. (18), with the aid of Eqs. (19)-(23), becomes
\begin{align}
V_{in}({\bf{r}}_{\parallel}, q; \omega)=\frac{2e^2}{\epsilon_b}\sum_{nn'}\,\sum_{mm'}\,\sum_{tt'}\,
                                        & \Pi_{nn'mm'tt'}(k, k'; \omega)\nonumber\\
&\times\!\!\! \int \!d{\bf{r}'}_{\parallel}\,
        K_0(q\vert{\bf{r}}_{\parallel}-{\bf{r}'}_{\parallel}\vert)\,
        \phi^*_{nm}({\bf{r}'}_{\parallel})\,\phi_{n'm'}({\bf{r}'}_{\parallel})\nonumber\\
&\times \big< n'm'\big\vert V_{tot}({\bf{r}''}_{\parallel}, q; \omega)\big\vert nm\big>\,S_{tt'}(q).
\end{align}
Now, we take matrix elements of both sides of this equation within the states $\big\vert i j\big>$ and
$\big\vert i'j'\big>$ to write
\begin{align}
\big<ij\big\vert V_{in}({\bf{r}}_{\parallel}, q; \omega)\big\vert i'j' \big>
=\sum_{nn'}\,\sum_{mm'}\,\sum_{tt'}\,&\Pi_{nn'mm'tt'}(k, k'; \omega)
 \, U_{ij,i'j',nm,n'm'}(q)\, S_{tt'}(q)\nonumber\\
&\, \times \big< n'm'\big\vert V_{tot}({\bf{r}''}_{\parallel}, q; \omega)\big\vert nm\big>,
\end{align}
where the matrix elements of the Coulombic interactions
\begin{align}
U_{ij,i'j',nm,n'm'}(q)
 =\frac{2e^2}{\epsilon_b}\int \!d{\bf{r}}_{\parallel}\int \!d{\bf{r}'}_{\parallel}\,
     \phi^*_{ij}({\bf{r}}_{\parallel})\,\phi_{i'j'}({\bf{r}}_{\parallel})
 K_0(q\vert{\bf{r}}_{\parallel}-{\bf{r}'}_{\parallel}\vert)
 \phi^*_{n'm'}({\bf{r}'}_{\parallel})\,\phi_{nm}({\bf{r}'}_{\parallel}).
\end{align}
Before we proceed further, let us define a few composite indices such as $\mu \equiv n, m$; $\mu'\equiv n', m'$;
$\nu \equiv i, j$; and $\nu' \equiv i', j'$ in order to rewrite Eq. (25) in the form
\begin{align}
\big<\nu\big\vert V_{in}({\bf{r}}_{\parallel}, q; \omega)\big\vert \nu'\big>
=\sum_{\mu \mu'}\,\sum_{tt'}\,\Pi_{\mu \mu' t t'}(k, k'; \omega)
U_{\nu \nu' \mu \mu'}(q)\, S_{tt'}(q)
\big<\mu'\big\vert V_{tot}({\bf{r}''}_{\parallel}, q; \omega)\big\vert \mu\big>,
\end{align}
where
\begin{align}
U_{\nu \nu' \mu \mu'}(q)=\frac{2e^2}{\epsilon_b}\,\int \!d{\bf{r}}_{\parallel}&\int \!d{\bf{r}'}_{\parallel}\,
     \phi^*_{\nu}({\bf{r}}_{\parallel})\,\phi_{\nu'}({\bf{r}}_{\parallel})
\,K_0(q\vert{\bf{r}}_{\parallel}-{\bf{r}'}_{\parallel}\vert)
\,\phi^*_{\mu'}({\bf{r}'}_{\parallel})\,\phi_{\mu}({\bf{r}'}_{\parallel}).
\end{align}
Now, we invoke the condition of self-consistency -- $V_{tot}=V_{ex}+V_{in}$ -- to rewrite Eq. (27)
\begin{align}
\big<\nu\big\vert V_{ex}({\bf{r}}_{\parallel}, q; \omega)\big\vert \nu'\big>
=\sum_{\mu \mu'}\,\Big[\delta_{\mu\nu}\delta_{\mu'\nu'} 
 - &\sum_{tt'}\,\Pi_{\mu \mu' t t'}(k, k'; \omega)
 U_{\nu \nu' \mu \mu'}(q)\, S_{tt'}(q)\Big]\nonumber\\
&\times \big<\mu'\big\vert V_{tot}({\bf{r}''}_{\parallel}, q; \omega)\big\vert \mu\big>.
\end{align}
Here $\delta_{\mu\nu}$ is the Kronecker delta. Since the external potential and the total potential are correlated
through the nonlocal, dynamic dielectric function $\epsilon$(...) such that

\begin{align}
V_{ex}({\bf{r}}_{\parallel}, q; \omega)=\int d{\bf{r}'}_{\parallel}\,
\epsilon({\bf{r}}_{\parallel},{\bf{r}'}_{\parallel}, q; \omega)\,
V_{tot}({\bf{r}'}_{\parallel}, q; \omega).
\end{align}
By taking the matrix elements of the left-hand side of this equation between the states $|\nu>$ and $|\nu'>$ and
adjusting a few terms, we obtain
\begin{align}
\big<\nu \big | V_{ex}(...)\big|\nu'\big>
&=\sum_{\mu}\sum_{\mu'}\,\big<\mu\nu\big|\epsilon (...)\big|\nu'\mu'\big>\,
                                                     \big<\mu'\big|V_{tot}(...)\big|\mu\big>\nonumber\\
&=\sum_{\mu\mu'}\,\epsilon_{\mu\nu\mu'\nu'}(...)\,\big<\mu'\big|V_{tot}(...)\big|\mu\big>.
\end{align}
A piecewise comparison of Eqs. (29) and (31) leaves us with the following:
\begin{align}
\epsilon_{\mu\nu\mu'\nu'}(q, \omega)
=\delta_{\mu\nu}\delta_{\mu'\nu'} - A_{\mu \mu'}(q, \omega)\,U_{\nu \nu' \mu \mu'}(q)\, ,
\end{align}
where
\begin{align}
A_{\mu\mu'}(q,\omega)=\sum_{tt'}\,\Pi_{\mu\mu'}(k,k'=k+q; \omega)\cdot S_{tt'}(q)
\end{align}
Equation (32) represents the generalized (nonlocal, dynamic) dielectric function for the system subjected to a lateral
confining harmonic potential and an applied magnetic field in the symmetric gauge. This is the main result, which can
be exploited to study a whole host of electronic, optical, and transport phenomena in the system. However, the
generalized dielectric function needs to be specified in order to make an otherwise impractical ($\infty \times \infty$)
matrix -- generated from Eq. (32) -- docile.
\begin{figure}[htbp]
\includegraphics*[width=8cm,height=14cm]{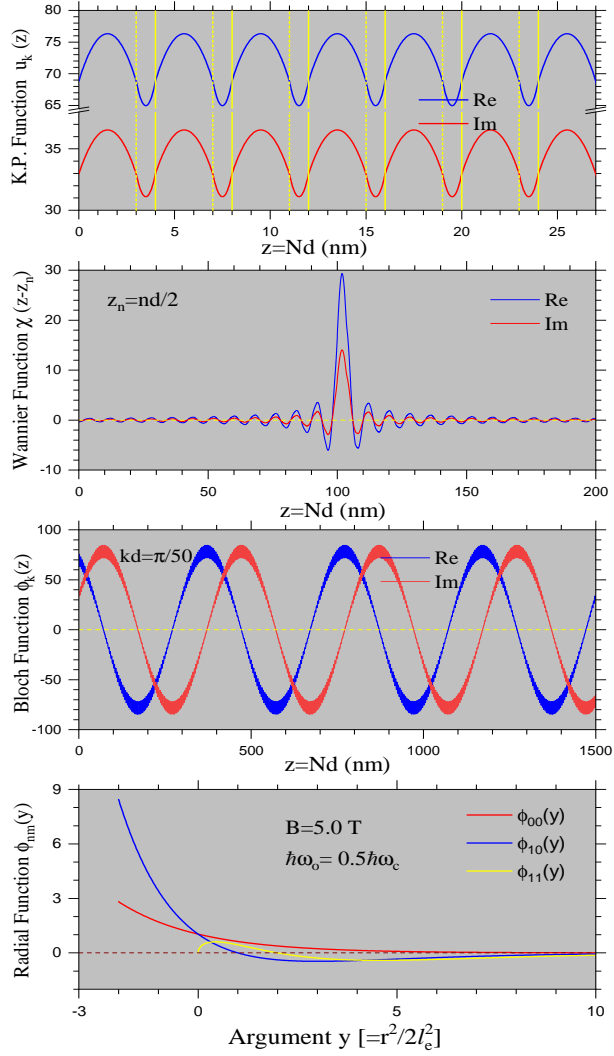}
\caption{(Color online) Graphic behavior of the KP, the Wannier, the Bloch, and the Laguerre functions as used in
this work. The well (barrier) thickness is 3 (1) nm and $V_0=349.11$ meV. The other parameters are as listed in
the respective panels.}
\label{fig3}
\end{figure}

\subsection{The KP, Wannier, and Bloch functions}

Let us briefly recall how the Wannier, Bloch, and KP functions are so tightly interwoven. The Wannier function of a
band $\rchi_t (z - z_n)$ is expressed in terms of the Bloch function $\phi_k(z)$ of the same band such that
\begin{align}
\rchi_t (z - z_n)=\frac{1}{\sqrt{N}}\,\sum_k\, e^{-i k z_n}\,\phi_k(z)\, ,
\end{align}
where the Bloch function $\phi_k(z)$ is defined in terms of the KP function $u(z)$ as
\begin{align}
\phi_k(z)=e^{i k z}\,u(z)\, ,
\end{align}
where $u(z)=u(z+d)$ satisfies the Schrodinger equation
\begin{align}
\Big[-\frac{\hbar^2}{2m^*}\,\frac{d^2}{d z^2} + V_c(z)\Big]\,u(z)=\epsilon_z\,u(z).
\end{align}
The Wannier function $\rchi_t(...)$ in Eq. (34) -- with the aide of Eq. (35) -- assumes the form
\begin{align}
\rchi_t (z - z_n)
=\frac{1}{\sqrt{N}}\,\sum_k\, e^{ik(z - z_n)}\,u(z)
=\sqrt{N}\,u(z)\,\frac{\sin[\pi(z-z_n)/d]}{[\pi(z-z_n)/d]}.
\end{align}
The Wannier functions and the Bloch functions satisfy the orthonormalization conditions just as explained in the text
following Eq. (8) above. Since the general (total) eigenfunction characterizing the system is expressed in terms of
these functions, we choose to illustrate their individual behavior [see Fig. 3]. As each panel reaffirms the exact
behavior of the respective function, we safely avoid expanding on the discussion since {\em a picture is worth a
thousand words}.


\subsection{The functional strategy: Specification}

As stated above, the generalized dielectric function (GDF) requires certain specifications to be met before it can be
exploited to obtain desired results such as the excitation spectrum of the system. First, we would like to underline
the z-motion along the superlattice axis (and the associated factors), which -- due, in fact, to the very nature of
the Lorentz force -- remains unaffected by the applied magnetic field ($B$) because ${\bf B}\parallel {\hat z}$. The
immediate relevance of the z-motion in the present system involves, for example, the miniband structure, the bandwidth,
and the overlap factor of $S_{tt'}(q)$. Let us limit ourselves {\em once and for all} to the lowest miniband structure
-- i.e., $t=0=t'$ -- for the sake of simplicity. In order to understand the miniband structure we make use of the
well-known Bastard's boundary conditions (BBC) because the active and inactive layers in the 1D superlattice system are
made up of different host materials [80]. Notice that we have already studied the details of the miniband structure (due
to the KP potential model) and the behavior of the overlap factor $S_{00}(q)$ in the zero-field case [62] and hence do
not feel an urge to expand on them here. Similarly, we had also discussed there the critical connection between the
symmetry (of the confining 2D harmonic potential) and the degeneracy in sufficient detail.

Now, we move on to discuss the necessary specification in the case of a non-zero magnetic field. Given that most of the
experiments on the low-dimensional systems are performed at lower temperatures, we limit ourselves to the absolute zero
(i.e., $T=0$ K). In this regard, we believe that the temperature dependence of our results would be significant only at
$T\gtrsim 35$ K. The absolute zero allows us to replace the Fermi distribution function with the Heaviside unit step
function such that
\begin{equation}
f(\epsilon)=\theta(\epsilon_F - \epsilon)=\left \{
\begin{array}{c}
1 \,\,\,\,\, {\rm if \,\,\,\,\, \epsilon_F > \epsilon}\\
0 \,\,\,\,\, {\rm if \,\,\,\,\, \epsilon_F < \epsilon}
\end{array}
\right . ,
\end{equation}
where $\epsilon_F$ is the Fermi energy in the system. The next important point is the subband ocupancy, which is implicit
in the suffix $\mu$, $\mu'$, $\nu$, and $\nu'$ in Eq. (32), for example. The GDF in Eq. (32) represents, in general, an $\infty\times\infty$ matrix until and unless we delimit the number of subbands and hence the electronic transitions to be
considered in the problem. While experiments may report multiple subbands occupied, theoretically it is vastly difficult
to compute the excitation spectrum for the multiple-subband model -- i.e., when $n$ ($-n<m<+n$) is a very big number. The
real reason is that the GDF turns out to be a matrix of the dimension of $\eta^2\times\eta^2$, where $\eta$ is the number
of subbands accounted for in the problem. Tackling such enormous matrices (for a large $\eta$) analytically is a {\em
hard nut to crack}, which is why we choose to confine ourselves to a two-subband model ($n,\,n'\equiv$ 1, 2) with only
the lowest one occupied and keep the complexity to a minimum. This is quite a sensible assumption for the small-density,
low-dimensional systems at lower temperatures where most of the experiments are performed.
As such, the GDF in Eq. (32) can be cast in the form:
\begin{align}
\tilde{\epsilon}(q,\omega)=
\left [
\begin{array}{rrrr}
1-B^{mm'jj'}_{0000} \ & \  -B^{mm'jj'}_{0001} \ & \  -B^{mm'jj'}_{0010} \ & \  -B^{mm'jj'}_{0011} \\
 -B^{mm'jj'}_{0100} \ & \ 1-B^{mm'jj'}_{0101} \ & \  -B^{mm'jj'}_{0110} \ & \  -B^{mm'jj'}_{0111} \\
 -B^{mm'jj'}_{1000} \ & \  -B^{mm'jj'}_{1001} \ & \ 1-B^{mm'jj'}_{1010} \ & \  -B^{mm'jj'}_{1011} \\
 -B^{mm'jj'}_{1100} \ & \  -B^{mm'jj'}_{1101} \ & \  -B^{mm'jj'}_{1110} \ & \ 1-B^{mm'jj'}_{1111} \\
\end{array}
\right ]\, ,
\end{align}
where $B^{mm'jj'}_{nn'ii'}=A^{mm'}_{nn'}\,U^{mm'jj'}_{nn'ii'}$ and $A^{mm'}_{nn'}=\Pi^{mm'}_{nn'}(k,k'=k+q;\omega)\cdot
S_{00}(q)$. Since subband index $n$ can take the value zero, we have specified $n, n' \equiv 0$, 1 -- rather than
$n, n'\equiv1$, 2. The superscripts refer to the orbital quantum numbers, which are defined such that $-n< m <+n$.
Finally, it is imperative to realize that the symmetry of the confining {\em harmonic} potential in the system brings in
some persuasive consequences. For a symmetric potential well, $U_{ijkl}$ (the Fourier-transformed Coulomb interaction)
is strictly zero for an arbitrary value of $q$ if the sum $i+j+k+l$ is an {\em odd} number. This is because the
analogous eigenfunction is either symmetric or antisymmetric under space reflection. This scenario is perfectly clear in
the zero-field case ($B=0$). For a finite $B$, the situation takes twists and turns due to the fact that the center of
the harmonic oscillator is displaced from zero due to the cyclotron orbits by a finite distance on the order of $x_c$ ($=q\ell^2_e$). However, a heedful mathematical manipulation of the integrands is all that is merely required to prove
that $U_{ijkl}$ is virtually zero even when $B\ne0$, provided that the sum $i+j+k+l$ is odd. The unoccupancy of the
second subband and the symmetry of the confining potential thus allow the matrix in Eq. (39) to assume the form
\begin{align}
\tilde{\epsilon}(q,\omega)&=\Big(1-B_{0000}^{0000}\Big)\nonumber\\
&\times \left [
\begin{array}{rrrrrr}
1-B_{0101}^{0-10-1} \ & \  -B_{0101}^{0-100} \ & \  -B_{0101}^{0-101} \ & \
 -B_{0110}^{0-1-10} \ & \  -B_{0110}^{0-100} \ & \  -B_{0110}^{0-110} \\
 -B_{0101}^{0+00-1} \ & \ 1-B_{0101}^{0+000} \ & \  -B_{0101}^{0+001} \ & \
 -B_{0110}^{0+0-10} \ & \  -B_{0110}^{0+000} \ & \  -B_{0110}^{0+010} \\
 -B_{0101}^{0+10-1} \ & \  -B_{0101}^{0+100} \ & \ 1-B_{0101}^{0+101} \ & \
 -B_{0110}^{0+1-10} \ & \  -B_{0110}^{0+100} \ & \  -B_{0110}^{0+110} \\
 -B_{1001}^{-100-1} \ & \  -B_{1001}^{-1000} \ & \  -B_{1001}^{-1001} \ & \
1-B_{1010}^{-10-10} \ & \  -B_{1010}^{-1000} \ & \  -B_{1010}^{-1010} \\
 -B_{1001}^{+000-1} \ & \  -B_{1001}^{+0000} \ & \  -B_{1001}^{+0001} \ & \
 -B_{1010}^{+00-10} \ & \ 1-B_{1010}^{+0000} \ & \  -B_{1010}^{+0010} \\
 -B_{1001}^{+100-1} \ & \  -B_{1001}^{+1000} \ & \  -B_{1001}^{+1001} \ & \
 -B_{1010}^{+10-10} \ & \  -B_{1010}^{+1000} \ & \ 1-B_{1010}^{+1010} \\
\end{array}
\right ]
\end{align}
A word of warning: the superscripts of the elements merely refer to the plus or minus of the respective indices without
using a comma in between (for brevity). The same plus or minus in front of some zeros are added only for the sake of
{\em spatial} symmetry of the elements. Now, we simplify the matrix in Eq. (40) by following the usual rules of matrix
arithmetic. So, let us subtract the 3rd column from the 1st and the 6th column from the 4th to write the resulting
matrix, which is further simplified by subtracting 3rd row from the 1st and the 6th row from the 4th to finally write
\begin{align}
\tilde{\epsilon}(q,\omega)&=\Big(1-B_{0000}^{0000}\Big)\nonumber\\
&\times \left [
\begin{array}{rrrrrr}
 2 \ & \ 0 \ & \  -1 \ & \ 0 \ & \  0 \ & \  0 \\
 0 \ & \ 1-B_{0101}^{0+000} \ & \  -B_{0101}^{0+001} \ & \  0 \ & \ -B_{0110}^{0+000} \ & \  -B_{0110}^{0+010} \\
-1 \ & \  -B_{0101}^{0+100} \ & \ 1-B_{0101}^{0+101} \ & \  0 \ & \ -B_{0110}^{0+100} \ & \  -B_{0110}^{0+110} \\
 0 \ & \ 0 \ & \  0 \ & \ 2 \ & \  0 \ & \  -1 \\
 0 \ & \  -B_{1001}^{+0000} \ & \  -B_{1001}^{+0001} \ & \  0 \ & \ 1-B_{1010}^{+0000} \ & \  -B_{1010}^{+0010} \\
 0 \ & \  -B_{1001}^{+1000} \ & \  -B_{1001}^{+1001} \ & \ -1 \ & \  -B_{1010}^{+1000} \ & \ 1-B_{1010}^{+1010} \\
\end{array}
\right ]
\end{align}
In obtaining Eq. (41), the double operations (as stated above) have also taken advantage of the fact that the terms of
the form of $A^{0,-1}_{01}[U^{0,-1,0,-1}_{0101}-U^{0,-1,0,1}_{0101}]$ vanish. This is because the square brackets yield
the integrands of the Coulomb matrix elements, which differ only in the signs of the sine terms, make the integrands odd,
and hence become zero. Since the collective excitations in a system are obtained by searching the zeros of the determinant
of $\tilde{\epsilon}(...)$, we obtain
\begin{align}
\Big(1-B_{0000}^{0000}\Big)\,
\Big[1-\Big(B_{0101}^{0000} + B_{1010}^{0000}\Big) - \big[2\big]\,\Big(B_{0101}^{0101} + B_{1010}^{1010}\Big)\Big]=0
\end{align}
after the straightforward but lengthy mathematical steps. Another appealing and engaging way of writing Eq. (42) is as
follows:
\begin{align}
\big(1-A_{00}^{00}\,U_{0000}^{0000}\big)\,
                    \Big[1 - C_{01}^{00}\,U_{0101}^{0000} - \big[2\big]\,C_{01}^{01}\,U_{0101}^{0101}\Big]=0
\end{align}
where
\begin{align}
A_{00}^{00}=\Pi^{00}_{00}\,S_{00}(q)\, , \\
C_{01}^{00}=A_{01}^{00}+A_{10}^{00}=\rchi^{00}_{01}\,S_{00}(q)\, ,\\
C_{01}^{01}=A_{01}^{01}+A_{10}^{10}=\rchi^{01}_{01}\,S_{00}(q)\, ,
\end{align}
where $\rchi^{00}_{01}=\Pi^{00}_{01}+\Pi^{00}_{10}$ and $\rchi^{01}_{01}=\Pi^{01}_{01}+\Pi^{01}_{10}$. Notice that
$\Pi^{00}_{00}$ and $\rchi^{mm'}_{01}$ refer, respectively, to the intrasubband and intersubband electronic transitions.
Equation (43) is actually the one which is treated at the computational level for all practical purposes. It is
interesting to note that in spite of the substantial complexity brought about by the applied magnetic field, the
intasubband and the intersubband excitations making up the complete excitation spectrum of the system are observed to be
clearly decoupled. The explicit expressions of the matrix elements of the Coulombic interactions $U^{mm'jj'}_{nn'ii'}$
required in Eq. (43) will be listed in the next section.

\subsection{The Long wavelength limit}

Given their importance, it is interesting to study the long wavelength limit (LWL) of the polarizability functions
such as $\Pi^{00}_{00}$, $\rchi^{00}_{01}$, and $\rchi^{01}_{01}$ involved in the process. To that end, the required
mathematical diagnosis is quite complex and lengthy, which is why we choose to outline the procedure briefly and
thereby restrain the length of the article. We begin with writing $\Pi^{mm'}_{nn'}$ at absolute zero, replace the sum
over $k$ with an integral -- having the upper (lower) limit as $k_F$ (-$k_F$) --  using the proper replacement rule,
solve the integral with the help of $\mathcal{x}\, 2.558.4$ on page 183 in GR (1994) in Ref. [79], simplify the result
of the integral using $\tanh^{-1}(x) = \frac{1}{2}\, \ln \big[\frac{1+x}{1-x}\big]$, and then finally writing the
series expansion of $\ln (1+x)$ in the limit of $x\ll 1$. Next, we already know that $\rchi^{mm'}_{nn'}=\Pi^{mm'}_{nn'}
+ \Pi^{m'm}_{n'n}$, so it becomes less cumbersome to write directly $\rchi^{mm'}_{nn'}$ for the specific values of the
suffixes. The result is
\begin{align}
\Pi^{00}_{00}&\simeq\,\frac{n_{1D}\, q^2}{m^*\, \omega^2} + O(q^2), \hspace{0.5cm} {\rm for\,\,\, n=0=n'\,\,\,
and \,\,\,\, m=0=m' }\, ,
\end{align}
where $m^*\,\big[=\hbar^2/(w_h d^2);\, w_h=W_0/2\big]$, for the intrasubband excitations and
\begin{align}
\rchi^{00}_{01}&\simeq\,\frac{2\,n_{1D}\,\epsilon_{00}}{\big[(\hbar\omega)^2 - (\epsilon_{00})^2\big]} + O(q),
      \hspace{0.5cm} {\rm \epsilon_{00}=\hbar\Omega;\hspace{0.5cm} for\,\,\, m=0=m' } \\
\rchi^{01}_{01}&\simeq\,\frac{2\,n_{1D}\,\epsilon_{01}}{\big[(\hbar\omega)^2 - (\epsilon_{01})^2\big]} + O(q),
  \hspace{0.5cm} {\rm \epsilon_{01}=-\tfrac{3}{2}\,\hbar\Omega-\tfrac{1}{2}\,\hbar\omega_c; \hspace{0.5cm}
                 for\,\,\,m=0,\,m'=+1 }
\end{align}
for the intersubband excitations. Equations (47)$-$(49) clearly attest the fact that the long wavelength limits of the polarizability functions are independent of the virtual dimensionality of the system and have the same forms for higher
dimensions [1].

\subsection{The density of states}

The knowledge of the density of states (DOS) is of paramount significance in many areas of physics and helps us explain
numerous classical and quantal phenomena in a given system. For instance, the zero, vanishingly low, and the high DOS
indicate, respectively, the gap, the psuedo-gap, and the free propagation of the respective wave in the system. This
implies that the computation of the DOS can (and does) help us in studying the band-gap engineering in a system. The
photonic and phononic crystals are the recent examples where the band-gap engineering played a crucial role in defining
their specific use and in making them two of the hottest topics in physics during the past three decades. The DOS -- $D(\epsilon)$ -- in the systems of diminishing dimensions is in essence the number of distinct states at a certain energy
level that electrons are allowed to fill and is one such inceptive feature of a system that clearly demonstrates the
dimensional dependence. The standard analytical diagnosis divulges that $D(\epsilon) \propto \epsilon^{1/2}$, $D(\epsilon) \propto \epsilon^{0}$, and $D(\epsilon)\propto \epsilon^{-1/2}$, respectively, in the 3D, 2D, and 1D systems. In the
extreme case of quantum confinement -- such as quantum dot systems -- the DOS is known to be $\delta-$function-like
inferring to the vanishing of thermal broadening [1]. We have derived the DOS for the present system on the basis of the single-particle energy in Eq. (7) and the result is
\begin{align}
D(\epsilon)
= \frac{2}{\pi\, d}\,\sum_{nm}\, \Big[w^2_h -\big(\epsilon-\epsilon_{nmt}\big)^2\Big]^{-1/2}\cdot
\theta \Big(w^2_h -\big(\epsilon-\epsilon_{nmt}\big)^2 \Big)
\end{align}
where $\theta(...)$ is the Heaviside step function and the symbol $\epsilon_{nmt}$ is defined as
\begin{align}
\epsilon_{nmt}=\big[n+\tfrac{1}{2}(\alpha+1)\big]\hbar\Omega + \tfrac{1}{2}m\hbar\omega_c + \epsilon_t
\end{align}

\begin{figure}[htbp]
\includegraphics*[width=8cm,height=9cm]{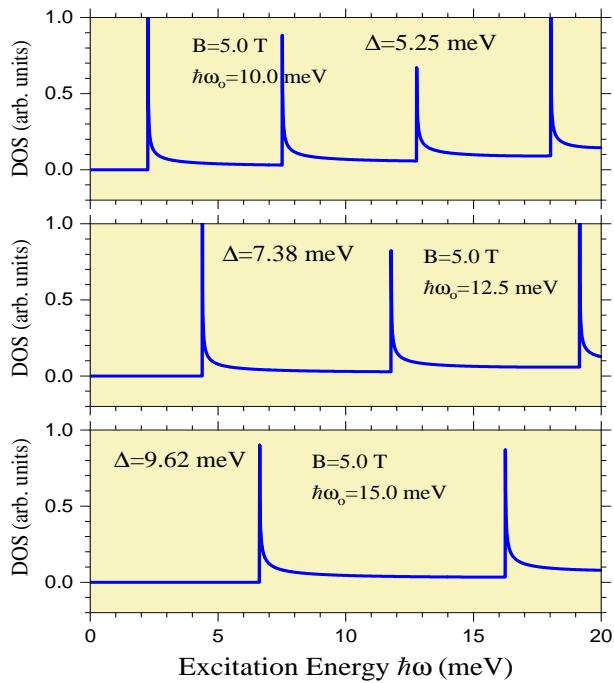}
\caption{(Color online) The density of states vs. the excitation energy for the VSQD system, for a magnetic field
of $B=5.0$ T. The confinement potential is defined as $\hbar \omega_o=10.0$ meV (top panel), 12.5 meV (middle
panel), and 15.0 meV (bottom panel). The band-width is $W_0=19.76$ meV.}
\label{fig4}
\end{figure}

Figure 4 shows the computed DOS versus the excitation energy for a given value of the magnetic field. Since the magnetic
field is constant, it is only the confinement potential that influences the shifting of the spikes in the three panels.
We observe that the DOS spikes start isolating from each other with increasing confinement potential. This process
clearly results in diminishing the number of spikes within a given energy range. In the simpler situations, it is the
subband spacing that defines the gap between the two consecutive spikes in the DOS. But, it is not the case with the
present periodic system of the VSQD. It is evident from three panels that the number of DOS spikes reduce with
increasing confinement. This tendency is found to obey a simple mathematical rule:
$n_f ={\rm Int}[n_i\times(\Omega_i/\Omega_f)]$, where $n_j$ is the number of peaks and $\Omega_j$ is the effective
characteristic frequency of the harmonic oscillator; $j\equiv i, f$. To be specific, the subindex $i(f)$ refers to the
initial (final) case of the history. This empirical rule was discovered in the case of a realistic quantum wire where it
remains unfailingly true both with and without an applied magnetic field [1].

\begin{figure}[htbp]
\includegraphics*[width=8cm,height=9cm]{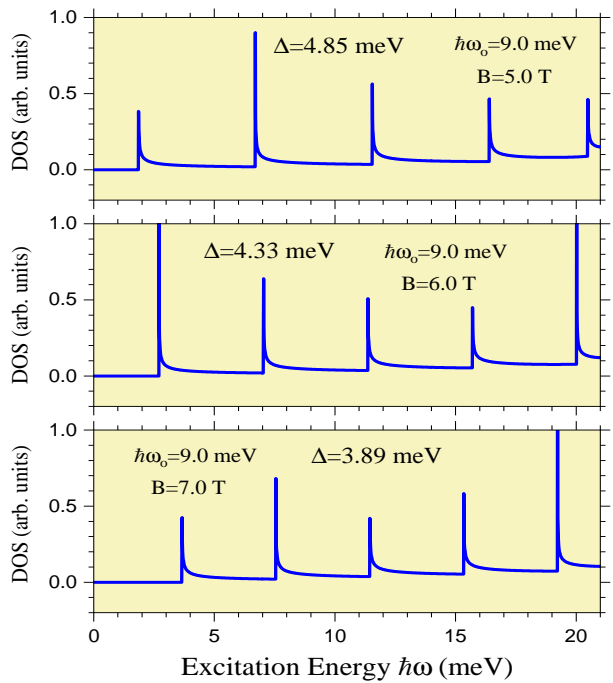}
\caption{(Color online) The density of states vs. the excitation energy for the VSQD system, for a confinement
potential of $\hbar\omega_o=9.0$ meV. The magnetic field is defined as $B=5.0$ T (top panel), 6.0 T
(middle panel), and 7.0 T (bottom panel). The band-width is $W_0=19.76$ meV. (After Kushwaha, Ref. 85).}
\label{fig5}
\end{figure}

Figure 5 shows the density of states versus the excitation energy for a given value of the confinement potential.
The magnetic field is defined as $B=5.0$ meV (top panel), 6.0 meV (middle panel), and 7.0 meV (bottom panel). We
observe that the DOS peaks start shifting towards the higher energy, but the energy separation between them
becomes smaller with increasing magnetic field. For $B=0$, it was found that while the DOS peaks shift towards
higher energy, they start distancing from each other with increasing confinement potential [62]. This implies
that the role of an applied magnetic field is virtually more than just to boost the confinement. In the case of
realistic quantum wires, we have noticed similar behavior regarding the peaks’ shifting, but the DOS peaks
experience larger energy separation with increasing magnetic field [1]. The empirical rule deduced in relation
with Fig. 4 does not hold good in this case (for the set of parameters used), whereas it remains valid for the
realistic quantum wires (both for $B=0$ and $B\ne0$) [1]. We think that the moderate confinement – with $B\ne0$ –
favors the VSQD in mimicking a realistic quantum wire.

\begin{figure}[htbp]
\includegraphics*[width=8cm,height=9cm]{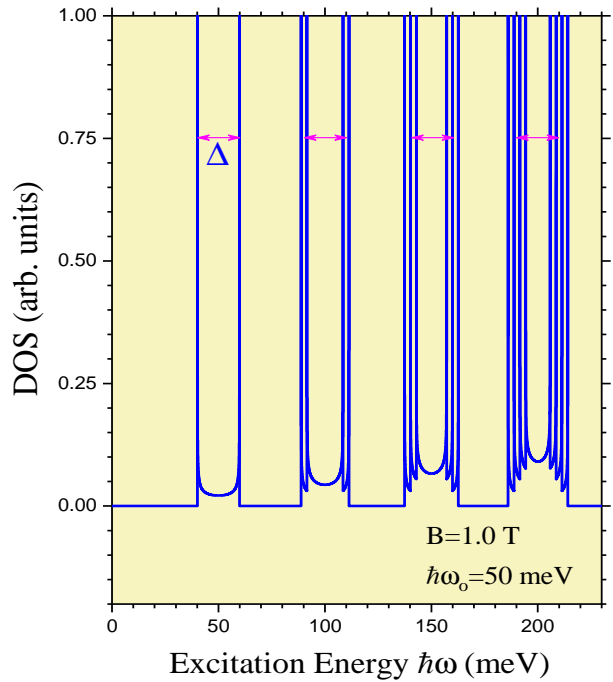}
\caption{(Color online) The density of states vs. the excitation energy for the VSQD system, for a magnetic field
of $B=1.0$ T and the confinement potential $\hbar\omega_o=50.0$ meV. The band-width is $W_0=19.76$ meV. The double
headed arrows (in magenta) refer to the average width $\Delta=W_0$.}
\label{fig6}
\end{figure}

Figure 6 displays the DOS as a function of the excitation energy for a typical magnetic field but an extremely strong
confinement potential. A too strong lateral (2D) confinement seems to bring the system of VSQD to a standstill by
causing disruption and thereby decoupling the InAs islands holding the quantum dots. This implies that the system in
this situation is clearly not anticipated to mimic the quantum wire. Interestingly, every (still broadened) spike is
seen to be made up of $2n'$ spikes and centered at the confinement harmonics ($n'.\hbar\omega_o$). This is
intelligible: given that $\omega_o \gg \omega_c$, Eq. (51) simplifies to $\epsilon_{nmt}=n'\cdot \hbar\omega_o$,
with $n'\equiv2$[$(n+\frac{1}{2}(|m|+1)$]; which implies that the average width of the spikes at every harmonic is
given by $\Delta=W_0$. This is exactly what we observe in the picture (see the double-headed arrows in magenta).

\subsection{The Fermi energy}

The textbooks on Condensed Matter Physics teach us the terms like Fermi gas, Fermi energy, Fermi level, Fermi surface,
Fermi temperature, and Fermi velocity, which are so closely associated with the spin-$1/2$ particles called fermions.
The Fermions are governed by the Pauli exclusion principle and they obey the Fermi-Dirac statistics. A precise
knowledge of these terms and of the distinction between them is vital to the understanding of electron dynamics of a
quantum system. We would like to highlight three of them: Fermi surface, which is defined as the Fermi-energy surface
in the reciprocal space; Fermi energy, which is the energy difference between the highest and lowest occupied states
in a system of non-interacting fermions at absolute zero; and Fermi level, which is the total energy level (including
kinetic and potential energies) that remains well-defined even in the complex interacting system at thermodynamic
equilibrium. Here, we are interested in the Fermi energy since we deal with the system that lies at absolute zero. The
Fermi energy derived on the basis of the single-particle energy in Eq. (7) is expressed as
\begin{align}
n_{1D} = \frac{2}{\pi\,d}\,\sum_{nm}\, \cos^{-1}\Big[-\frac{1}{w_h}\,\big(\epsilon_F - \epsilon_{nmt}\big)\Big]
\cdot \theta \Big[-\frac{1}{w_h}\,\big(\epsilon_F - \epsilon_{nmt}\big) \Big]
\end{align}
where $n_{1D}$ is the linear charge density of the electrons in the system. This is exactly the formula we make use
of to compute the Fermi energy for various instances.

\begin{figure}[htbp]
\includegraphics*[width=7cm,height=9cm]{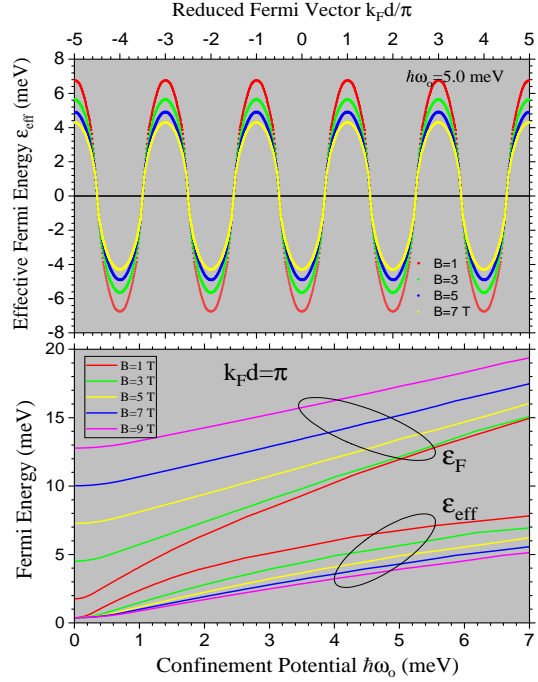}
\caption{(Color online) Upper panel: The effective Fermi energy vs. the reduced Fermi vector, for confinement
potential $\hbar \omega_o=5.0$ meV and for several values of the magnetic field. We purposely made this choice
to have $k_F d$ on the abscissa -- it could very well be the charge density $n_{1D}$. Lower panel: The Fermi
energy vs. the confinement potential, for a specific value of $k_F$ and for several values of $B$. The lower
(upper) set of curves refers to the effective Fermi (Fermi) energy in the system. The band-width $W_0=19.76$ meV.}
\label{fig7}
\end{figure}

The upper panel of Figure 7 illustrates the {\em effective} Fermi energy as a function of the reduced Fermi vector
($k_F d/\pi$) for a given value of the confinement potential $\hbar \omega_o$ and for several values of the magnetic
field $B$. The effective Fermi energy $\epsilon_{eff}$ is defined as
$\epsilon_{eff}=\epsilon_{F} -\tfrac{1}{2}\,\hbar\Omega$. Notice that $\epsilon_{eff}$ oscillates as a function of
the Fermi vector $k_F$, for any value of magnetic field. This oscillatory behavior is an outright reflection of the
cosine term in the single-particle energy due to the tight-binding approximation [see Eq. (7)]. Interestingly, the
peak-to-peak amplitude of $\epsilon_{eff}$ decreases with increasing $B$. This is clearly comprehensible and follows
the fact that the cyclotron radius $r_c \propto B^{-1}$. The abscissa in Fig. 7 can also account for the linear
charge density in the system, since the Fermi vector is related to the 1D charge density by a simple relation:
$k_F=(\pi/2) n_{1D}$.
The lower panel shows the Fermi energy $\epsilon_{F}$ as well as the effective Fermi energy $\epsilon_{eff}$ as a
function of the confinement potential $\hbar \omega_o$, for a given value of the Fermi vector $k_F$ and for several
values of the magnetic field $B$. The lower panel can very well be considered complementary to the upper one as it
substantiates the trend of $\epsilon_{eff}$ in the latter for any value of $\hbar \omega_o$. The knowledge of the
Fermi energy is central to the understanding of nearly all electronic, optical, and transport properties of a quantal
system at a given temperature. The transport properties -- such as conductance or resistance -- are true reflections
of the electron dynamics at the Fermi surface. What is so captivating about the Fermi surface is that it can be
tailored before it tailors the rest in the system.

\begin{figure}[htbp]
\includegraphics*[width=7cm,height=9cm]{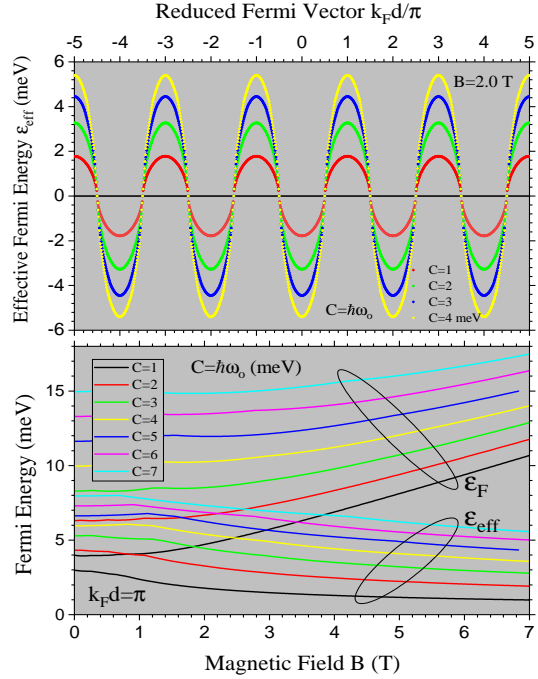}
\caption{(Color online) Upper panel: The effective Fermi energy vs. the reduced Fermi vector, for the magnetic field
$B=2$ T and for several values of the confinement potential. Lower panel: The Fermi energy vs. the magneticfield,
for a specific value of $k_F$ and for several values of $\hbar\omega_o$. The lower (upper) set of curves refers to
the effective Fermi (Fermi) energy in the system. The band-width $W_0=19.76$ meV.}
\label{fig8}
\end{figure}

The upper panel of Figure 8 exemplifies the {\em effective} Fermi energy as a function of the reduced Fermi vector
($k_F d/\pi$) for a given value of the magnetic field $B$ and for several values of the confinement potential
$\hbar\omega_o$. The effective Fermi energy $\epsilon_{eff}$ is observed to oscillate as a function of the Fermi v
ector $k_F$ for any value of $\hbar\omega_o$. Again, this oscillatory behavior is attributed to the cosine term in
the single-particle energy in Eq. (7). But, in contrast to Fig. 7, the peak-to-peak amplitude of the effective Fermi
energy is observed to increase with increasing confinement. This is the observation we expect intuitively for a
constant -- and even for zero [62] -- magnetic field. The lower panel shows the Fermi energy as well as the effective
Fermi energy versus the magnetic field, for a given value of the Fermi vector $k_F$ and for several values of the
confinement potential. Again, the lower set of curves representing $\epsilon_{eff}$ substantiates our observation in
the upper panel. Comparing Figs. 7 and 8 leads us to infer that while the Fermi energy increases with increasing both
magnetic field and confinement, the effective Fermi energy decreases (increases) with increasing magnetic field
(confinement) -- while keeping the confinement (magnetic field) constant.

\subsection{The Coulombic interactions}

The matrix elements of the Coulombic interactions, which inherently give the legitimacy to the formation of the
collective excitations in the system, defined in Eq. (26) [or (28)] are expressed in the simplified form as
follows:
\begin{align}
U^{mm'jj'}_{nn'ii'}(q)
 =\frac{2\,e^2}{\epsilon_b}\,\frac{1}{(2\pi)^2}\,
 \sqrt{\frac{n!}{(n+\vert m\vert)!}}\,
 \sqrt{\frac{n'!}{(n'+\vert m'\vert)!}}\,
 \sqrt{\frac{i!}{(i+\vert j\vert)!}}\,
 \sqrt{\frac{i'!}{(i'+\vert j'\vert)!}}\,\nonumber\\
 \times \int dy \int dy'\, e^{-y}\,e^{-y'}\,
 y^{\vert m\vert/2}\,y^{\vert m'\vert/2}\,y'^{\vert j\vert/2}\,y'^{\vert j'\vert/2}\nonumber\\
 \times L^{\vert m\vert}_n(y)\,L^{\vert m'\vert}_{n'}(y)\,L^{\vert j\vert}_i(y')\,
                                                         L^{\vert j'\vert}_{i'}(y')\nonumber\\
 \times \int d\theta \,\int d\theta'\,
 e^{-i(m-m')\theta}\, K_{o}\Big(q_r\big\vert {y}^2+{y'}^2-2yy'\cos(\theta-\theta')\big\vert^{1/2}\Big)\,
                      e^{i(j-j')\theta'}\,.
\end{align}
where $q_r=\sqrt{2}q\ell_e$ and $y$ and $y'$ are dimensionless variables. We have made use of the relation between
the confluent hypergeometric function and the generalized Laguerre polynomial [79]
\begin{align}
L^{(\alpha)}_n (y)=\binom {n+\alpha}{n}\,\Phi(-n, \alpha+1, y)\, ,
\end{align}

\begin{figure}[htbp]
\includegraphics*[width=8cm,height=9cm]{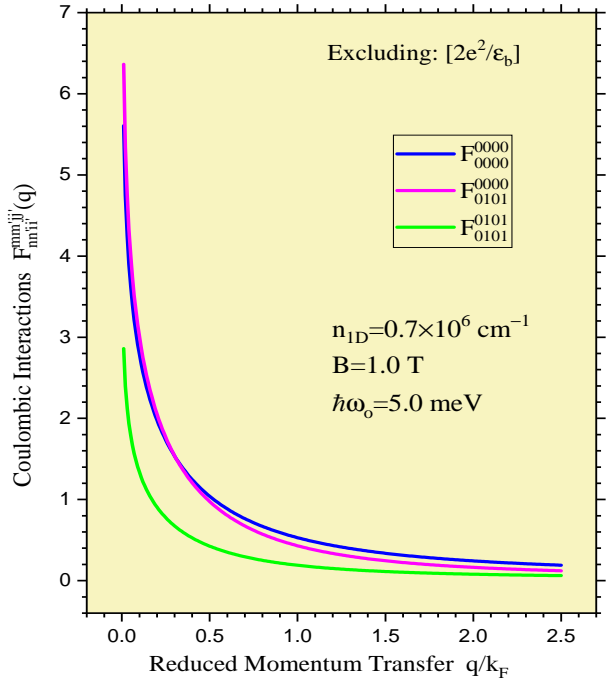}
\caption{(Color online) The Fourier-transformed Coulombic interactions $F^{0000}_{0000}(q)$, $F^{0000}_{0101}(q)$,
and $F^{0101}_{0101}(q)$ plotted as a function of the reduced momentum transfer $q/k_F$. We call attention to the $F^{0000}_{0101}(q)$ being slightly dominant over the $F^{0000}_{0000}(q)$ in the range $0\le q/k_F \le 0.31$ [see
the text]. (After Kushwaha, Ref. 85).}
\label{fig9}
\end{figure}

Figure 9 manifests the computed matrix elements of the Fourier-transformed Coulombic interactions $F^{mm'nn'}_{ii'jj'}$,
where the superscripts (subscripts) refer to the orbital (principal) quantum number. The non-zeroness of these elements
grants electronic excitations a many-particle character because of which the respective excitations are termed
collective excitations: specifically, they are called plasmons (magnetoplasmons) depending upon whether the magnetic
field is zero (nonzero) in the system. The collective excitations become Landau-damped after merging with the respective
single-particle continuum in the short wavelength limit (SWL). We notice that
$F^{0000}_{0000}(q) > F^{0000}_{0101}(q) > F^{0101}_{0101}(q)$ over a greater part of the momentum transfer $q$, except
for the range of $0<q/k_F<0.31$ within which $F^{0000}_{0101}(q)$ is slightly greater than $F^{0000}_{0000}(q)$. Such a
singular range is no trivial matter to the excitation spectrum and can (and does) have valid consequences. Concisely,
within $0<q/k_F<0.31$, the system fosters a metastable state, which is associated with the corresponding magnetoplasmon
owning a negative group velocity (NGV). The attestation of NGV signals anomalous dispersion in a gain medium with the
population inversion, which is the basis for the lasing action of lasers [81, 82].

\section{Illustrative Examples on the magnetoplasmons}

For the illustrative numerical examples, we focus on the InAs/GaAs system just as in the original experiments [4-6]. The material parameters used are [85]: effective mass $m^*=0.042 m_{_0}$ ($0.067 m_{_0}$) and the background dielectric
constant $\epsilon_{_b}=13.9$ ($12.8$) for the InAs (GaAs). We employ the potential barrier of height $V_0=349.11$ meV
that produces the bandwidth (of the lowest ($t=0$) miniband) $W_0=19.76$ meV, in compliance with Sakaki [2] so as to
minimize the optical phonon scattering. The confinement potential $\hbar\omega_{_0}=5.0$ meV, the effective Fermi energy $\epsilon_{eff}=6.4238$ meV for a 1D charge density $n_{1D}=0.7 \times10^6$ cm$^{-1}$, and the effective width of the
confining (parabolic) potential well, estimated as the FWHM of the extent of the eigenfunction, $w_{eff}=2\sqrt{2 \ln (2)}\sqrt{n+1}\,\ell_{_c}=22.022$ nm. Note that the Fermi energy $\epsilon_F$ is dependent on the charge density
($n_{1D}$) and the confining potential ($\hbar \omega_{o}$). Thus, our goal is to investigate the single-particle and
collective excitations in a quantum wire comprised of the VSQD in a two-subband model within the full RPA [76] at
absolute zero (T=0 K). The success of RPA in such diverse geometries as the electrically/magnetically modulated systems
[83] and the Rashba spintronic systems [84] is nothing but praiseworthy.

\subsection{The influence of the layer widths}

In this section, we discuss the principal results of this investigation on the excitation spectrum of the magnetized
VSQD. The main aim behind discussing these results is to see the consequential changes with the variation in the well
and barrier widths in the system. It has been found that the layer widths are the most intriguing parameters and seem
to drastically influence the results on the excitation spectrum. The influence of the magnetic field and/or the
confinement potential on the excitation spectrum will be discussed later in a separate section.

\begin{figure}[htbp]
\includegraphics*[width=8cm,height=9cm]{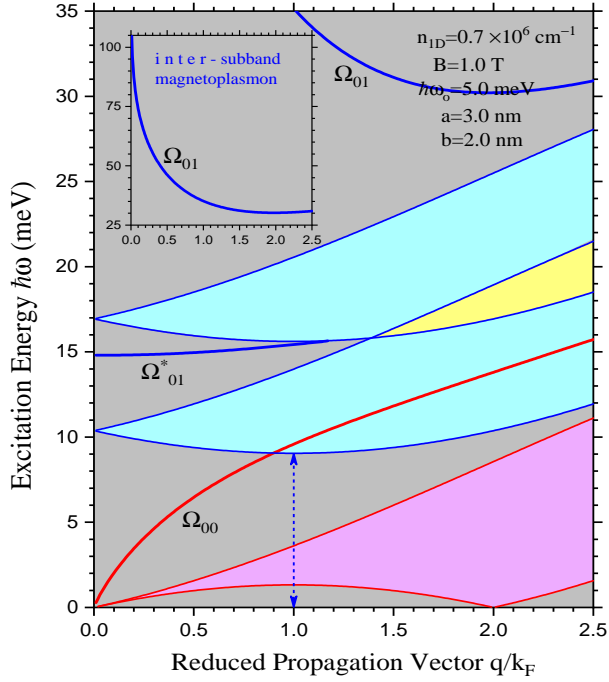}
\caption{(Color online) The excitation spectrum of the VSQD where the energy $\hbar \omega$ is plotted as a function
of the reduced momentum transfer $q/k_F$ in the situation where $a>b$.
The shaded part in magenta (cyan) refers to the intrasubband (intersubband) SPE associated with the lowest occupied
(first excited) subband at $T=0$. The bold red (blue) curve represents the intrasubband (intersubband) CME. The
vertical double-headed arrow stresses the minimum of the intersubband SPE at $q=k_F$. We call attention to gap-CME ($\Omega^*_{01}$) which starts from the origin and propagates to merge with the lower branch of the upper split
intersubband SPE before the point of intersection of the inner branches. The higher intersubband CME starts from zero
with NGV, observes a minimum, and then changes the sign of its group velocity before merging with the respective SPE.
All the relevant parameters are as listed inside the picture. (After Kushwaha, Ref. 85).}
\label{fig10}
\end{figure}

Figure 10 illustrates the full excitation spectrum of the resultant quantum wire comprised of the VSQD in a two-subband
model within the RPA, for the case where well-width $a$ is greater than the barrier width $b$. The full spectrum is
composed of the single-particle excitations (SPE) and collective (magnetoplasmon) excitations (CME) for a given set of parameters: $n_{1D}$, $\hbar\omega_o$, $B$, $a$, and $b$. The figure caption specifies the SPE and CME with all
essential details. The intrasubband CME ($\Omega_{00}$) starts from the origin and propagates without tending to merge
with the respective SPE up until $q/k_F=2.5$. The intersubband SPE bifurcates at the origin and an intersubband CME ($\Omega^*_{01}$) appears at $\hbar \omega=14.809$ meV to propagate within the gap and finally merge with the lower edge
of the upper split SPE (at $q/k_F=1.335$, $\hbar\omega=15.768$ meV), just before the intersection of the inner edges of
the SPE. The upper intersubband CME ($\Omega_{01}$) arises from the origin at $\hbar\omega=102.963$ meV with a NGV but
changes its sign at $q/k_F \simeq 2.01$ to thence propagate with a positive group velocity (PGV) until it merges with
the respective SPE in the SWL. The fascinating part of the excitation spectrum is that the leading CME do not merge with
the respective SPE, thus do not suffer from the Landau damping, and hence remain long-lived excitations over the greater
part of momentum transfer. These CME make their case effortless for the Raman scattering experiments. As to the gap-CME ($\Omega^*_{01}$), its origination is solely ascribed to the non-zero orbital quantum number $m$, it is equally bonafide,
and should become easily observable for all the right reasons. It is not difficult to justify the single-particle energy
at the critical points: $q/k_F =$ 0, 1, and 2. The energy difference between the SPE and the CME at the origin is a manifestation of the many-body effects such as the depolarization and excitonic shifts [1]. For the set of parameters
used, the mean radius of the quantum dots confined within the InAs islands is estimated
(via $R^2_o/{2\ell^2_e}=1$) to be $R_o\simeq19$ nm.

\begin{figure}[htbp]
\includegraphics*[width=8cm,height=9cm]{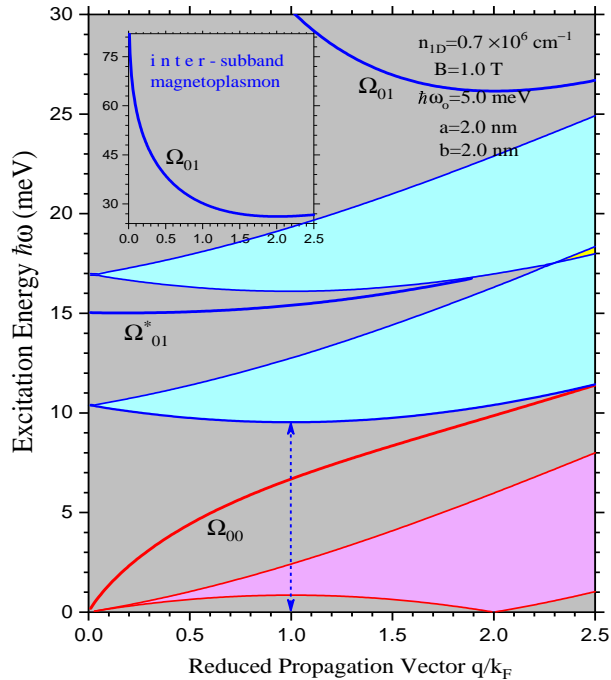}
\caption{(Color online) The same as in Fig. 10, but for the case where the well and the barrier widths are equal:
$a=b$. The excitation spectrum observes lowering of the whole spectrum in energy and smoothing out of the big
dip in the higher intersubband CME. We call attention to the point of intersection of the split intersubband SPE,
which has moved to the higher momentum transfer $q$ as compared to that in Fig. 10. Also, there is a wider gap
between the intrasubband and intersubband SPE continua than seen in Fig. 10. All the relevant parameters are as
listed inside the picture.}
\label{fig11}
\end{figure}

Figure 11 portrays the full excitation spectrum for the resultant system of VSQD in the situation in which well width
$a$ is equal to the barrier width $b$. A close comparison with Fig. 10 reveals that the whole excitation spectrum
lowers in energy, the point of intersection of the inner branches of the split intersubband SPE moves towards the
higher $q$, and a wide gap is opened up between the intrasubband and intersubband SPEs with decreasing well-width.
The intrasubband CME starts from the origin and propagates without tending to merge with the respective SPE up
until $q/k_F=2.5$. The intersubband SPE splits into two with a gap-CME ($\Omega^*_{01}$) starting from the origin
at $\hbar \omega=15.036$ meV and propagating within the gap to merge with the lower edge of the upper split SPE at
($q/k_F=1.893$, $\hbar\omega=16.804$ meV), just before the point of intersection of the inner edges of the SPE. The
upper intersubband CME now starts from the origin at $\hbar\omega=82.137$ meV with the NGV but changes its sign at
$q/k_F \simeq 2.01$ to propagate with PGV before it merges with the respective SPE in the short wavelength limit
(SWL). As to the gap-CME ($\Omega^*_{01}$), it remains equally bonafide, propagates up until a higher $q$ than in
Fig. 10, and should become easily observable in the Raman scattering experiments. The rest of the discussion related
with Fig. 10 is still valid.

\begin{figure}[htbp]
\includegraphics*[width=8cm,height=9cm]{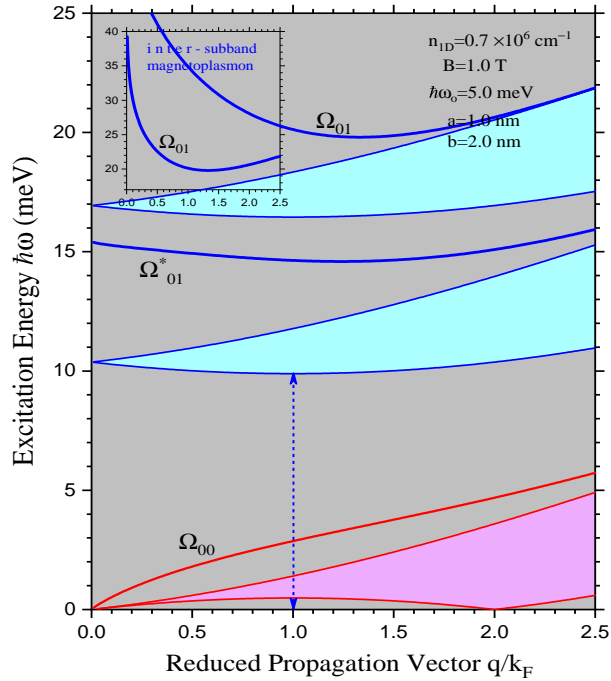}
\caption{(Color online) The same as in Fig. 10, but in the case where the well width is smaller than the barrier
width: $a<b$. We call attention to the still further lowering of the whole spectrum in energy and a complete gap
between the split intersubband SPE up until $q/k_F=2.5$. The gap-CME has now total legitimacy and is free from
the Landau damping all the way up to $q/k_F=2.5$. There is now a larger gap between the intersubband and
intrasubband SPE continua than in Fig. 11. An exact diagnosis verifies that the lower (upper) edge of the split
intersubband SPE in Figs. $10-12$ starts at 10.372 (16.936) meV.}
\label{fig12}
\end{figure}

Figure 12 shows the full excitation spectrum for the system of VSQD in the situation when the well width $a$
is smaller than the barrier width $b$. A close observation reveals that the whole excitation spectrum further lowers
in energy in comparison with Fig. 11, a full gap is now opened up between the split intersubband SPEs, and there is
a larger gap between the intrasubband and intersubband SPEs with decreasing well-width. The intrasubband CME starts
from the origin and propagates without tending to merge with the respective SPE up until $q/k_F=2.5$. The gap-CME ($\Omega^*_{01}$) starts from the origin at $\hbar \omega=15.405$ meV and propagates freely within the gap tending
to merge with the upper edge of the lower split SPE somewhere in the short wavelength limit (SWL). The upper
intersubband CME now starts from the origin at $\hbar\omega=39.254$ meV with the NGV but changes its sign
at $q/k_F \simeq 1.33$ to propagate with positive group velocity in the close vicinity of -- but without merging
with  -- the upper edge of the upper split intersubband SPE up until $q/k_F=2.5$. The rest of the discussion related
with Fig. 10 is still valid.

\subsection{The influence of the magnetic field}

The complete excitation spectrum of a system -- within a specified model -- is comprised of the single-particle and
the collective excitations plotted together. Interestingly, the role of the single-particle excitations in the full
spectrum is two-fold: (i) to define the propagation range of the collective excitations, and (ii) to specify whether
or not they are free from Landau damping within the given $\omega-q$ space. In other words, the single-particle
excitations specify the life of and provide the legitimacy to the collective excitations. However, it is up to the
bonafide collective excitations to profess their role in the device design -- for a given set of parameters.

\begin{figure}[htbp]
\includegraphics*[width=8cm,height=9cm]{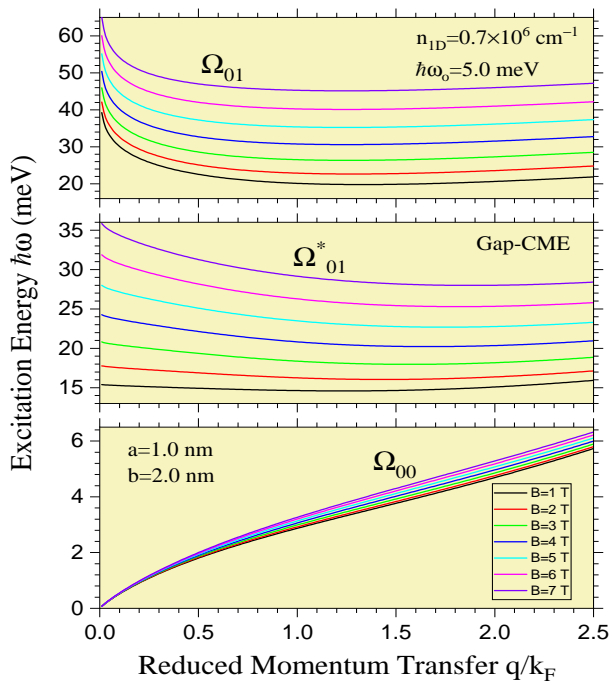}
\caption{(Color online) The collective (magnetoplasmon) excitation spectrum within a two-subband model where the
energy $\hbar \omega$ is plotted as a function of the reduced momentum transfer $q/k_F$, for several values of
the magnetic field $B$. The other parameters are as listed inside the picture. Notice that we have only plotted
the bonafide intrasubband and intersubband collective (magnetoplasmon) excitations, which remain Landau-undamped
until a very large momentum transfer.}
\label{fig13}
\end{figure}

Figure 13 displays the energy of the collective (magnetoplasmon) excitations versus the reduced momentum transfer for
a given confinement potential and for several values of the applied magnetic field. We choose the set of parameters
corresponding to that for Fig. 12. This implies three clear-cut CMEs: the intrasubband CME ($\Omega_{00}$), the
gap-CME ($\Omega^*_{01}$), and the highest CME ($\Omega_{01}$) which emerges from the origin with a NGV plotted,
respectively, in the lower, the middle, and the upper panels. As stated above, we have avoided to plot the respective single-particle excitations in order not to make a mess in the picture. What we observe in this figure is obviously
what we intuitively expect: the energy of the magnetoplasmon excitations increases with increasing magnetic field. It
is worth mentioning that this observation remains true independently of the size, shape, and dimensions of the system
--  with a very few exceptions. It is interesting to note that while $\Omega_{00}$ indicates a monotonous increase in
the energy, $\Omega^*_{01}$ shows a tendency to emerge with an NGV, and $\Omega_{01}$ generally maintains the trend
of originating with an NGV -- with increasing magnetic field. The gap-CME and the uppermost CME, which start with
an NGV, generally, observe a minimum before changing the sign of their group velocity and thence propagate with a PGV
until they merge with the respective SPE and become Landau-damped at the higher momentum transfer.

\subsection{On the inverse dielectric functions}

The diehard, traditional, condensed-matter theorists recognize that searching the zeros of the dielectric function and
the poles of the inverse dielectric function (IDF) must, in principle, produce identical results for the excitation
spectra in a system [78]. This is very true but not without certain reservations. The latter has for sure characteristic advantages over the former. For instance, the imaginary (real) part of the IDF furnishes a significant measure of the longitudinal (Hall) resistance in the system. This indicates that investigating IDF allows us to comprehend not only the
optical but also the transport phenomena in a system of interest. Besides, examining Im [$\epsilon^{-1}(q,\omega)$] also
tacitly provides precise details of the inelastic electron (or Raman) scattering cross-section in the system. In relation
to the inelastic electron scattering, Im [$\epsilon^{-1}(q,\omega)$] plays a central role in studying such remarkable
phenomena as the fast-particle energy loss to the nanostructure in question.

\begin{figure}[htbp]
\includegraphics*[width=8cm,height=9cm]{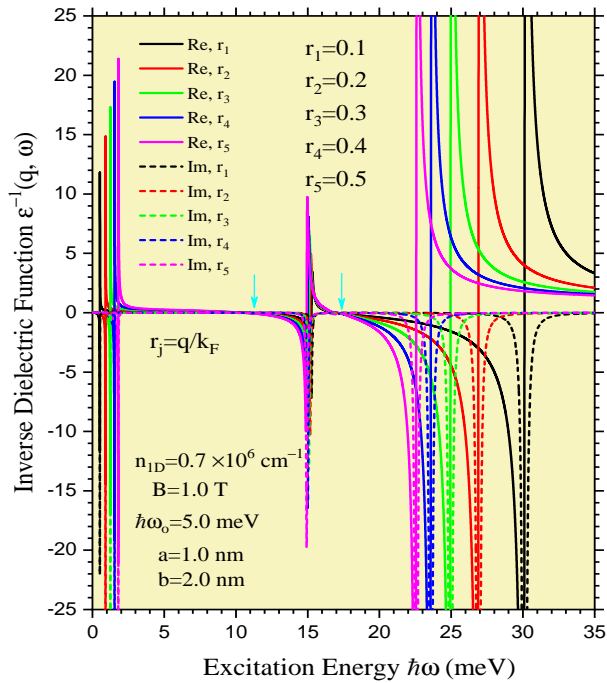}
\caption{(Color online) Inverse dielectric function $\epsilon^{-1}(q,\omega)$ vs. the excitation energy
$\hbar\omega$ for the given values of the momentum transfer $q/k_F$. The other parameters are listed in
the picture. (After Kushwaha, Ref. 85).}
\label{fig14}
\end{figure}

Figure 14 depicts the inverse dielectric function versus the excitation energy of the system, for various values of the
momentum transfer $q/k_F$. The real (imaginary) parts of the IDF are plotted as solid (dashed) curves for a given $q/k_F$.
Since we have handpicked relatively small values of $q/k_F$, we must anticipate reproducing the collective (rather than
the single-particle) excitations. It is verified that all three peaks, for a given $q/k_F$, correspond to the respective
CME in the excitation spectrum. A small energy span covered by the middle peaks -- pertaining to the gap-CME
($\Omega^*_{01}$) -- indicates that the gap-mode is pretty much flat in the long wavelength limit (LWL), which is very
true (see, e.g., Figs. 10-12). To be specific, the resonance peaks arising at
$\hbar\omega \,{\rm(meV)}=$ 0.519, 0.913, 1.247, 1.542, and 1.807, for example, reproduce the intrasubband CME in Fig. 12.
The resonance peaks occurring at $\hbar\omega \,{\rm(meV)}=$ 15.312, 15.217, 15.131, 14.961, and 14.885 yield the gap-CME
in Fig. 12. With the same token, the peaks at $\hbar\omega \,{\rm(meV)}=$ 30.018, 26.816, 24.876, 23.513, and 22.631
furnish the upper intersubband CME in Fig. 12. It is noteworthy that both the gap-CME and the upper CME are seen to exist
in the reverse order to the values of $q$. This is because both the gap-CME and the upper CME propagate with a NGV until
they observe a minimum. The details of the magnetorotonic character of the upper CME are deferred to a future publication.

\begin{figure}[htbp]
\includegraphics*[width=8cm,height=9cm]{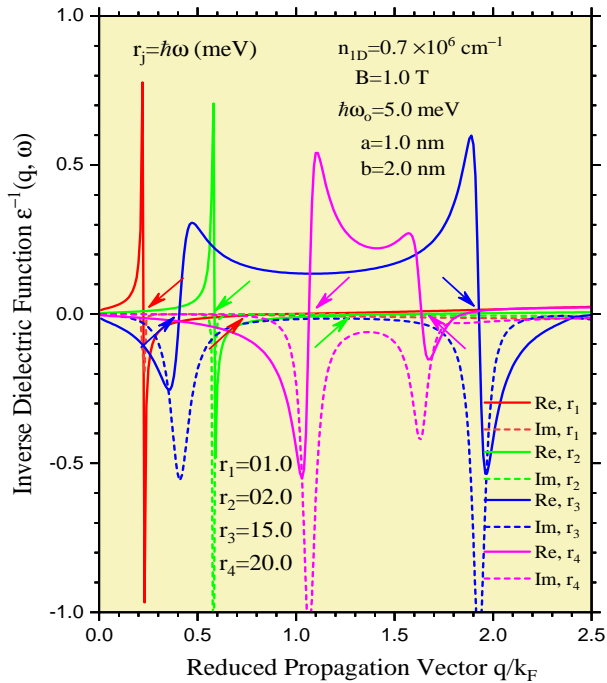}
\caption{(Color online) Inverse dielectric function $\epsilon^{-1}(q,\omega)$ vs. the momentum transfer
$q/k_F$ for the given values of the excitation energy $\hbar\omega$. The other parameters are as listed
inside the picture. A pair of identical colored arrows indicate the resonance peaks and the {\em crossings}
of the real and imaginary parts, which we assume must yield a good correspondence with Fig. 12.}
\label{fig15}
\end{figure}

Figure 15 instances the inverse dielectric function versus the momentum transfer of the system $q/k_F$, for several
values of the excitation energy $\hbar\omega$. The real (imaginary) parts of the IDF are plotted as solid (dashed)
curves for a given value of $\hbar\omega$. Needless to say the quantity that directly strikes the quantum transport
is the spectral weight Im[$\epsilon^{-1}(q,\omega)$] which controls both the single-particle contribution at large
momentum transfer ($q$) and the collective (magnetoplasmon) contribution at small $q$. Given the set of the
selected excitation energies $\hbar\omega=$ 1.0, 2.0, 15.0, and 20.0, we expect to have covered parts of all three
CME as well as the intrasubband SPE in Fig. 12. The pair of red and green arrows indicating the resonance peaks at
($q/k_F=0.226, \hbar\omega=1.0$) and ($q/k_F=0.582, \hbar\omega=2.0$) the crossings at
($q/k_F=0.763, \hbar\omega=1.0$) and ($q/k_F=1.305, \hbar\omega=2.0$) reproduce, respectively, the intrasubband CME
($\Omega_{00}$) and the SPE in Fig. 12. The pair of blue arrows indicating the resonance peaks at
($q/k_F=0.405, \hbar\omega=15.0$) and ($q/k_F=1.925, \hbar\omega=15.0$) reproduce the gap-CME
($\Omega^*_{01}$) in Fig. 12. Similarly, the pair of magenta arrows indicating the resonance peaks
at ($q/k_F=1.065, \hbar\omega=20.0$) and ($q/k_F=1.636, \hbar\omega=20.0$) reproduce the uppermost intersubband CME ($\Omega_{01}$) in Fig. 12. As such, Figs. 14 and 15 substantiate the fact that the zeros of the DF and the poles
of the IDF yield exactly identical results (see above). However, this is merely a glimpse of the efficacy of the
IDF. In order to commend fully the role of IDF, one has to dive deep down to learn their exact analytical diagnosis
as well as to pursue their practicality for a given system [78].

%
%

\section{Concluding Remarks}

In summary, the author has investigated the single-particle and collective (magnetoplasmon) excitations in a resultant
system of quantum wire made up of the VSQD subjected to a 2D (planar) confining (parabolic) potential and an applied
magnetic field in the symmetric gauge within a two-subband model in the framework of full RPA. The noteworthy features
of this report are the following: (i) the B-dependence of the DOS authenticating the claim of VSQD mimicking a
realistic quantum wire, (ii) the Fermi energy oscillating as a function of the Bloch vector, (iii) an unusual
bifurcating of the intersubband SPE at the origin owing to the non-zero B, (iv) the existence of a bonafide CME within
the gap of the split SPE, (v) all three CME – including the gap mode – being free from the Landau damping up until a
relatively large momentum transfer, (vi) the gap CME becoming better defined and remaining Landau-undamped even in the
SWL ($q/k_F \ge 2$) with decreasing period of the system, (vii) The creation of very very high-energy (upper) intersubband
CME in the LWL, (viii) the upper intersubband CME propagating largely with the NGV until a wave vector twice the Fermi
vector, (ix) the B-dependence of the IDF setting to furnish a significant measure of the transport phenomena in the
system. We have neglected the strain effect because its influence on the collective excitations – plasmons and/or magnetoplasmons – is practically nonexistent. Finally, the exact IDF as studied here knows no bounds regarding the
subband occupancy and serves a useful purpose of analyzing the inelastic electron (and Raman) scattering experiments.
In a nutshell, the small length scales and the strong coupling along the growth axis effectuate the resultant
nanostructure of VSQD to mimic a realistic quantum wire. This clearly emboldens our confidence in the theoretical
framework of the problem and brings our quest for {\em reversing the trend} to a fruitful finish. The size-dependence
of the tunability of the energy range achieved by the CME makes the periodic system of VSQD a prospective source of
offering magnetoplasmon qubits with a super high-speed advantage over the exciton qubits – particularly in the LWL [86].
Given the realistic set of geometric and material parameters – which are within the reach of the current nanotechnology – involved in the process, the realization of the periodic system of VSQD seems to be a fact rather than a fiction. We
believe that such behavior characteristics of the CME as investigated and predicted here can be successfully observed by
the resonant Raman scattering experiments. Consequently, the magnetized chain of VSQD must serve to be a potentially
viable system for implementing the formal idea of quantum state transfer for the quantum computation and the quantum
communication networks.

\begin{acknowledgments}
This work is a comprehensive version of a {\em Letter} recently published in EPL (see Ref. 85). The author has
benefited from stimulating discussions and communications with some outstanding colleagues in the field. I
would like to particularly thank Hiroyuki Sakaki, Allan MacDonald, Bahram Djafari-Rouhani, Peter Nordlander,
and Douglas Natelson to mention a few. I also wish to acknowledge Kevin Singh for the timely assistance with the
software during the course of this work.
\end{acknowledgments}
\newpage
\noindent {\bf Data Availability:} The  data  that  support  the  findings  of  this  study  are  available within
the article.






\begin{references}
\bibitem[1]{1} For an extensive review of electronic, optical, and transport phenomena in the systems of reduced
               dimensions, such as quantum wells, quantum wires, quantum dots, and  (electrically/magnetically)
               modulated systems, see M.S. Kushwaha, Surf. Sci. Rep. {\bf 41}, 1 (2001).
\bibitem[2]{2} H. Sakaki, Jpn. J. Appl. Phys. {\bf 28}, L314 (1989).
\bibitem[3]{3} B.Y.K. Hu and S. Das Sarma, Phys. Rev. Lett. {\bf 68}, 1750 (1992).
\bibitem[4]{4} Q. Xie, A. Madhukar, P. Chen, and N.P. Kobayashi, Phys. Rev. Lett. {\bf 75}, 2542 (1995).
\bibitem[5]{5} G.S. Solomon, J.A. Trezza, A.F. Marshall, and J.S. Harris, Phys. Rev. Lett. {\bf 76}, 952 (1996).
\bibitem[6]{6} M.S. Miller, Jpn. J. Appl. Phys. {\bf 36}, 4123 (1997).
\bibitem[7]{7} T. Inoue, M. Asada, N. Yasuoka, T. Kita, and O. Wada, J. Phys.: Conf. Series {\bf 245}, 012076 (2010).
\bibitem[8]{8} Y. Shoji, R. Oshima, A. Takata, and Y. Okada, J. Cryst. Growth {\bf 312}, 226 (2010).
\bibitem[9]{9} Ch. Huang, M. Igarashi, S. Horita, M. Takeguchi,Y. Uraoka, T. Fuyuki, I. Yamashita, and S. Samukawa,
               Jap. J. App. Phys. {\bf 49}, 04DL16 (2010).
\bibitem[10]{10} S.K. Zhang, Th. Myint, W.B. Wang, B.B. Das, N. Perez-Paz, H. Lu, M.C. Tamargo, A. Shen, and
               R. R. Alfano, J. Vac. Sci. Technol. B {\bf 28}, C3D17 (2010).
\bibitem[11]{11} J.J. Zhang, N. Hrauda, H. Groiss, A. Rastelli, J. Stangl, F. Schäffler, O.G. Schmidt, and G. Bauer,
               Appl. Phys. Lett. {\bf 96}, 193101 (2010).
\bibitem[12]{12} T. Inoue, M. Asada, N. Yasuoka, O. Kojima, T. Kita, O. Wada, Appl. Phys. Lett. {\bf 96},
                 211906 (2010).
\bibitem[13]{13} Y. Shoji, R. Oshima, A. Takata, and Y. Okada, Physica E {\bf 42}, 2768 (2010).
\bibitem[14]{14} G. Koblmuller, S. Hertenberger, K. Vizbaras, M. Bichler, F. Bao, J.P. Zhang, and G. Abstreiter,
                 Nanotechnology {\bf 21}, 365602 (2010).
\bibitem[15]{15} O. Moshe, D.H. Rich, B. Damilano, and J. Massies, J. Phys. D: Appl. Phys. {\bf 44}, 505101 (2011).
\bibitem[16]{16} K. Kuklinski, Ł. Kłopotowski, K. Fronc, P. Wojnar, T. Wojciechowski, M. Czapkiewicz, J. Kossut,
                 G. Karczewski and T. Wojtowicz, Acta Physica Polonica A {\bf 120}, 856 (2011).
\bibitem[17]{17} Y. Ikeuchi, T. Inoue, M. Asada, Y. Harada, T. Kita, E. Taguchi1, and H. Yasuda. Appl. Phys.
                 Express {\bf 4} 062001 (2011).
\bibitem[18]{18} W.S. Liu, H.M. Wu, Y.A. Liao, J.I. Chyi, W.Y. Chen , and T.M. Hsu, J. Cryst. Growth {\bf 323}, 164
                 (2011).
\bibitem[19]{19} K.N. Hui and K.S. Hui, Current Appl. Phys. {\bf 11}, 662 (2011).
\bibitem[20]{20} S.S. Walavalkar, A.P. Homyk, C.E. Hofmann, M.D. Henry, C. Shin, H.A. Atwater, and A. Scherer,
                 Appl. Phys. Lett. {\bf 98}, 153114 (2011).
\bibitem[21]{21} K. Akahane, N. Yamamoto, M. Naruse, T. Kawazoe, T. Yatsui, and M. Ohtsu, Jap. J. Appl. Phys.
                 {\bf 50}, 04DH05 (2011).
\bibitem[22]{22} D. Kim, S.G. Carter, A. Greilich, A.S. Bracker and D. Gammon, Nature Phys. {\bf 7}, 223 (2011).
\bibitem[23]{23} T.W. Saucer, J.E. Lee, A.J. Martin, D. Tien, J.M. Millunchick, and V. Sih, Solid State Commun.
                 {\bf 151}, 269 (2011).
\bibitem[24]{24} K. Muller, A. Bechtold, C. Ruppert, C. Hautmann, J. S. Wildmann, T. Kaldewey, M. Bichler,
                 H. J. Krenner, G. Abstreiter, M. Betz, and J. J. Finley, Phys. Rev. B {\bf 85}, 241306 (2012).
\bibitem[25]{25} D. Hauser, G. Savelli, M. Plissonnier, L. Montès, and J. Simon, Thin Solid Films {\bf 520}, 4259
                 (2012).
\bibitem[26]{26} E. Koroknay, W.M. Schulz, D. Richter, U. Rengstl, M. Reischle, M. Bommer, Ch. A. Kessler1, Robert
                 Roßbach, H. Schweizer, M. Jetter, and P. Michler, Phys. Stat. Solidi B {\bf 249}, 737 (2012).
\bibitem[27]{27} B. Diaz, A. Malachias, L.A. Montoro, P.H.O. Rappl, E. Abramof, Nanotechnology {\bf 23}, 015604
                 (2012).
\bibitem[28]{28} I.H. Chen, K.H. Chen, C.C. Wang, and P.W. Li, ECS Transactions {\bf 50}, 313 (2012).
\bibitem[29]{29} A. Greilich, S.C. Badescu, D. Kim, A.S. Bracker, and D. Gammon, Phys. Rev. Lett. {\bf 110}, 117402
                 (2013).
\bibitem[30]{30} T. Sugaya, R. Oshima, K. Matsubara, and S. Niki, J. Appl. Phys. {\bf 114}, 014303 (2013).
\bibitem[31]{31} A. Takahashi, T. Ueda, Y. Bessho, Y. Harada, and T. Kita, Phys. Rev. B {\bf 87}, 235323 (2013).
\bibitem[32]{32} K.G. Eyink, L.J. Bissell, J. Shoaf, D.H. Tomich, D. Esposito, M. Hill, L. Grazulis, A. Aronow,
                 and K. Mahalingam, J. Vac. Sci. Technol. {\bf 31}, 03C131 (2013).
\bibitem[33]{33} D.Y. Oh, S.H. Kim, J. Huang, A. Scofield, D. Huffaker, A. Scherer, Nanotechnology {\bf 24},
                 265201 (2013).
\bibitem[34]{34} Y. Shoji, K. Akimoto and Y. Okada, J. Phys. D: Appl. Phys. {\bf 46}, 024002 (2013).
\bibitem[35]{35} M. Suwa, A. Takahashi, T. Ueda, B. Yusuke, Y. Harada, T. Kita, Phys. Stat. Solidi C {\bf 10},
                 1492 (2013).
\bibitem[36]{36} A.I. Yakimov, V.V. Kirienko, V.A. Armbrister, A.A. Bloshkin, and A.V. Dvurechenskii, Phys. Rev.
                 B {\bf 90}, 035430 (2014).
\bibitem[37]{37} D. Sonnenberg, A. Küster, A. Graf, Ch. Heyn and W. Hansen, Nanotechnology {\bf 25}, 215602 (2014).
\bibitem[38]{38} J. Wu, Y. Hirono, X. Li , Z.M. Wang, J. Lee, M. Benamara, S. Luo, Y.I. Mazur, E.S. Kim, and G.J.
                 Salamo, Adv. Funct. Mater. {\bf 24}, 530 (2014).
\bibitem[39]{39} V. Tasco, M. Usman, M.D. Giorgi and A. Passaseo, Nanotechnology {\bf 25}, 055207 (2014).
\bibitem[40]{40} V. Lopes-Oliveira, Y.I. Mazur, L.D. de Souza, L.A.B. Marcal, J. Wu, M.D. Teodoro, A. Malachias,
                 V.G. Dorogan, M. Benamara, G.G. Tarasov, E. Marega, G.E. Marques, Zh. M. Wang, M. Orlita, G.J.
                 Salamo, and V. Lopez-Richard, Phys. Rev. B {\bf 90}, 125315 (2014).
\bibitem[41]{41} Yu.I. Mazur, V. Lopes-Oliveira, L.D. de Souza, V. Lopez-Richard, M.D. Teodoro, V.G. Dorogan, M.
                 Benamara, J. Wu, G.G. Tarasov, E. Marega, Z. M. Wang, G.E. Marques, and G.J. Salamo, J. Appl.
                 Phys. {\bf 117}, 154307 (2015).
\bibitem[42]{42} J.H. Park, A. Mandal, D.Y. Um, S. Kang, D. Lee, and C.R. Lee, RSC Advances {\bf 5}, 47090 (2015).
\bibitem[43]{43} C.M. Chow, A.M. Ross, D. Kim, D. Gammon, A.S. Bracker, L.J. Sham, and D.G. Steel, Phys. Rev. Lett.
                 {\bf 117}, 077403 (2016).
\bibitem[44]{44} M.H. Kuo, S.K. Chou, Y.W. Pan, S.D. Lin, T. George, and P.W. Li, J. Appl. Phys. {\bf 120}, 233106
                 (2016).
\bibitem[45]{45} Y.H. Roh, S.J. Sim, I.J. Cho, N. Choi, and K.W. Bong, Analyst {\bf 141}, 4578 (2016).
\bibitem[46]{46} A. Küster, Ch. Heyn, A. Ungeheuer, G. Juska, S.T. Moroni2, E. Pelucchi and W. Hansen, Nanoscale
                 Res. Lett. {\bf 11}, 282 (2016).
\bibitem[47]{47} W. Xu, W. Liu, J.F. Schmidt, W. Zhao, X. Lu, T. Raab, C. Diederichs, W. Gao, D.V. Seletskiy and Q.
                 Xiong, Nature {\bf 541}, 62 (Jan 5, 2017).
\bibitem[48]{48} Y. Zhang, K.G. Eyink, L. Grazulis, M. Hill, J. Peoples, K. Mahaling, J. Cryst. Growth
                 {\bf 477}, 19 (2017).
\bibitem[49]{49} D.H. Yeon, B.C. Mohanty, C.Y. Lee, S.M. Lee, and Y.S. Cho, ACS Omega {\bf 2}, 4894 (2017).
\bibitem[50]{50} X. Tang, F. Wu, and King W.Ch. Lai, Appl. Phys. Lett. {\bf 110}, 241104 (2017).
\bibitem[51]{51} X. Song, Y. Zhang, H. Zhang, Y. Yu, M. Cao, Y. Che, H. Dai, J. Yang, X. Ding, and J. Yao,
                 Nanotechnology {\bf 28}, 145201 (2017).
\bibitem[52]{52} R. Long, D. Casanova, W.H. Fang, and O.V. Prezhdo, J. Am. Chem. Soc. {\bf 139}, 2619 (2017).
\bibitem[53]{53} A. Mohanta, D.J. Jang, S.K. Lu, D.Ch. Ling, and J.S. Wang, Appl. Phys. Lett. {\bf 110}, 033107
                 (2017).
\bibitem[54]{54} I. Seker, A. Karatutlu, O. Gurbuz, S. Yanik, Y. Bakis, and M. Karakiz, Appl. Phys. A {\bf 124},
                 47 (2018)
\bibitem[55]{55} S. Blumenthal, T. Rieger, D. Meertens, A. Pawlis, D. Reuter, and D.J. As, Phys. Stat. Solidi B
                 {\bf 255}, 1600729 (2018).
\bibitem[56]{56} A. Lin, M.F. Doty, and G.W. Bryant, Phys. Rev. B {\bf 99}, 075308 (2019).
\bibitem[57]{57} H.Y. Ramirez and Shun-Jen Cheng, Phys. Rev. Lett. {\bf 104}, 206402 (2010).
\bibitem[58]{58} A.I. Yakimov, A.A. Bloshkin, and A.V. Dvurechenskii, Phys. Rev. B {\bf 81}, 115434 (2010).
\bibitem[59]{59} K. Gawarecki, M. Pochwała, A. Grodecka–Grad, and P. Machnikowski, Phys. Rev. B {\bf 81},
                 245312 (2010).
\bibitem[60]{60} I. Mondragon-Shem, B.A. Rodríguez, and F.E. López, Comput. Phys. Commun. {\bf 181}, 1510 (2010).
\bibitem[61]{61} S. Prabhakar and R. Melnik, J. Appl. Phys. {\bf 108}, 064330 (2010).
\bibitem[62]{62} M.S. Kushwaha, J. Chem. Phys. {\bf 135}, 124704 (2011); and references therein.
\bibitem[63]{63} K. Gawarecki and P. Machnikowski, Acta. Phys. Pol. {\bf 119}, 637 (2011).
\bibitem[64]{64} A. Sitek and P. Machnikowski, Phys. Rev. B {\bf 86}, 205315 (2012).
\bibitem[65]{65} M. Usman, Phys. Rev. B. {\bf 86}, 155444 (2012).
\bibitem[66]{66} Li-Bo Chen, L.J. Sham, E. Waks, Phys. Rev. B. {\bf 85}, 115319 (2012).
\bibitem[67]{67} W.J. Pasek, M.P. Nowak, and B. Szafran, Phys. Rev. B. {\bf 89}, 245303 (2014).
\bibitem[68]{68} Li-Bo Chen and W. Yang, Laser Phys. Lett. {\bf 11}, 105201 (2014).
\bibitem[69]{69} K. Gawarecki, P. Machnikowski, and T. Kuhn, Phys. Rev. B. {\bf 90}, 085437 (2014).
\bibitem[70]{70} P. Karwat, and P. Machnikowski, Phys. Rev. B. {\bf 91}, 125428 (2015).
\bibitem[71]{71} T. Kawazu, Jap. J. Appl. Phys. {\bf 54}, 04DJ01 (2015).
\bibitem[72]{72} C. Segarra, J.I. Climente, F. Rajadell, and J. Planelles, J. Phys.: Condens. Matter {\bf 27},
                 415301 (2015).
\bibitem[73]{73} X. Ma, G.W. Bryant, and M.F. Doty, Phys. Rev. B. {\bf 93}, 245402 (2016).
\bibitem[74]{74} J.R. Jarzynka, P.G. MacDonald, J. Shumway, and I. Galbraith, J. Appl. Phys. {\bf 119}, 224303
                 (2016).
\bibitem[75]{75} P. Karwat, K. Gawarecki, and P. Machnikowski, Phys. Rev. B. {\bf 95}, 235421 (2017).
\bibitem[76]{76} B. Ilahi, K. Alshehria, N.A. Madhar, L. Sfaxi, H. Maaref, Results in Phys. {\bf 9}, 904 (2018).
\bibitem[77]{77} D. Pines, {\it The Many-Body Problem} (Benjamin, New York, 1961); A.L. Fetter and J.D. Walecka,
                 {\it Quantum Theory of Many-Particle Systems} (McGraw-Hill, New York, 1971); G.D. Mahan,
                 {\it Many Particle Physics} (Plenum, New York, 1981).
\bibitem[78]{78} M.S. Kushwaha, AIP Advances {\bf 2}, 032104 (2012); {\bf 3}, 042103 (2013); {\bf 6}, 035014 (2016);
                 {\bf 4}, 127151 (2014). These offer the exact derivations of the inverse dielectric functions (IDF)
                 for the Q-2D, Q-1D, and Q-0D electron systems of current interest and their ability to interpret
                 the two catholic experiments -- inelastic electron scattering and inelastic light (or Raman)
                 scattering -- in condensed matter physics.
\bibitem[79]{79} M. Abramowitz and I. A. Stegun, {\em Handbook of Mathematical Functions} (Dover, New York, 1972);
                 J. Spanier and K. B. Oldham, {\em An Atlas of Functions} (Springer-Verlag, Berlin, 1987); I.S.
                 Gradshteyn and I.M. Ryzhik, {\em Tables of Integrals, Series, and Products} (Academic, New York,
                 1994).
\bibitem[80]{80} G. Bastard, Phys. Rev. B {\bf 24}, 5693 (1981). The so-called ``Bastard's boundary conditions"
                 were known in the literature much before Bastard used them. See, e.g., D.J. BenDaniel and
                 C.B. Duke, Phys. Rev. {\bf 152}, 683 (1966).
\bibitem[81]{81} M.S. Kushwaha, Phys. Rev. B {\bf 78}, 153306 (2008); J. Appl. Phys. {\bf 109}, 106102 (2011);
                 Mod. Phys. Lett. B {\bf 28}, 1430013 (2014).
\bibitem[82]{82} M.S. Kushwaha, Europhys. Lett. {\bf 123}, 34001 (2018); Mod. Phys. Lett. B {\bf 33}, 1950062
                 (2019); This contains an Appendix on the gauge invariance.
\bibitem[83]{83} M.S. Kushwaha and H. Sakaki, Phys. Rev. B {\bf 69}, 155331 (2004); Solid State Commun. {\bf 130},
                 717 (2004).
\bibitem[84]{84} M.S. Kushwaha, Phys. Rev. B {\bf 74}, 045304 (2006); {\bf 76}, 245315 (2007); J. Appl. Phys.
                 {\bf 104}, 083714 (2008).
\bibitem[85]{85} M.S. Kushwaha, Europhys. Lett. {\bf 127} 37004 (2019).
\bibitem[86]{86} The precise diagnosis of the thematic observations such as the negative group velocity, the
                 magnetorotonic character, and the greater-speed magnetoplasmon qubits associated with the highest magnetoplasmons (in Figs. 10-12) is beyond the scope of this work and hence deferred to a future
                 publication.

\end{references}
\end{document}